\documentclass[11pt,fleqn]{article}
\usepackage{amssymb}
\usepackage{epsfig}
\usepackage{lscape,graphics,amsmath,geometry}
\usepackage{rotate}                                                     
\newcommand {\beq} {\begin{eqnarray}}
\newcommand {\eeq} {\end{eqnarray}}

\newcommand {\exclusion} [1] {}

\newcommand{\GeV} {\mathrm{GeV}}
\newcommand{\TeV} {\mathrm{TeV}}
\newcommand{\fb}  {\mathrm{fb}}
\newcommand{\fbi} {\mathrm{fb}^{-1}}
\newcommand{\cL } {{\cal L}}
\newcommand{\cP } {{\cal P}}
\newcommand{\cB } {{\cal B}}
\def\ee{e^+e^-}
\def\ti    {\tilde}
\def\stau  {{\ti\tau}}
\def\st    {{\ti t}}
\def\sb    {{\ti b}}
\def\cx    {\ti {\chi}}
\def\nt    {\ti {\chi}^0}
\def \Eslash {E \kern-.75em\slash }

\def\lsim{\raise0.3ex\hbox{$\;<$\kern-0.75em\raise-1.1ex\hbox{$\sim\;$}}}
\def\gsim{\raise0.3ex\hbox{$\;>$\kern-0.75em\raise-1.1ex\hbox{$\sim\;$}}}

\usepackage{latexsym}
\usepackage[dvips]{color}
\def\greaterthansquiggle{\raise.3ex\hbox{$>$\kern-.75em\lower1ex\hbox{$\sim$}}}
\def\lessthansquiggle{\raise.3ex\hbox{$<$\kern-.75em\lower1ex\hbox{$\sim$}}}

\newcommand{\grts}{\greaterthansquiggle}

\newcommand{\la}{\label}

\newcommand{\ci}{\cite}
\newcommand{\beqn}{\begin{eqnarray}}
\newcommand{\eeqn}{\end{eqnarray}}
\newcommand{\bequ}{\begin{equation}}
\newcommand{\eequ}{\end{equation}}
\newcommand{\bsl}{\begin{sloppypar}}
\newcommand{\esl}{\end{sloppypar}}

\definecolor{Black}{named}{Black}
\definecolor{Blue}{named}{Blue}
\definecolor{Red}{named}{Red}
\definecolor{Green}{named}{ForestGreen}
\definecolor{Black}{named}{Black}
\definecolor{Olive}{named}{OliveGreen}
\definecolor{Royal}{named}{RoyalBlue}
\definecolor{Orange}{named}{YellowOrange}
\definecolor{Yellow}{named}{Goldenrod}
\definecolor{Cornblue}{named}{CornflowerBlue} 
\definecolor{Lila}{named}{DarkOrchid}
\begin{document}
%
\null
\hfill DCPT-03-04\\
\null
\hfill DESY 03-030\\
\null
\hfill IPPP-03-02\\
\null
\hfill hep-ph/0303110

\vskip .4cm

\begin{center}
{\Large\bf
Polarisation in Sfermion Decays:\\[.3em] 
Determining {\boldmath $\tan\beta$} and Trilinear Couplings}
\vskip 1.5em

{\large
{
E. Boos$^{a,b}$, 
H.-U. Martyn$^{c}$, 
G. Moortgat-Pick$^{b,d}$,
M. Sachwitz$^{e}$, 
A. Sherstnev$^{a}$, and 
P.M. Zerwas$^{b}$
}                     
}\\[3ex]

{\footnotesize \it 
\noindent
$^{a}$ Skobeltsyn Institute of Nuclear Physics, Moscow State University,
119992 Moscow, Russia\\
$^{b}$ DESY, Deutsches Elektronen-Synchrotron, D-22603 Hamburg, Germany \\
$^{c}$ Rheinisch-Westf\"alische Technische Hochschule, D-52074 Aachen, 
Germany \\
$^{d}$ IPPP, University of Durham, Durham DH1 3LE, United Kingdom  \\
$^{e}$ DESY, Deutsches Elektronen-Synchrotron, D-15738  Zeuthen, Germany \\
}
\end{center}
\vskip .5em
\par
\vskip .4cm

\begin{abstract} \noindent
The basic parameters of supersymmetric theories can be determined at
future $e^+e^-$ linear colliders with high precision. 
We investigate in this report how polarisation measurements 
in $\tilde{\tau}$ and $\tilde{t}$ or
$\tilde{b}$ decays to $\tau$ leptons and $t$ quarks 
plus neutralinos or charginos
 can be used to measure $\tan\beta$ (in
particular for large values) and 
to determine the trilinear couplings $A_{\tau}$,
$A_t$ and $A_b$ in sfermion pair production.
\end{abstract}

\section{Introduction}
If supersymmetry is realised in Nature \cite{All1, All2}, a large
number of low--energy parameters -- masses, couplings and mixings --
must be determined with high precision. This is necessary in order to
investigate the mechanism  breaking the symmetry and to reconstruct
the fundamental theory eventually at a scale close to the Planck scale
\cite{Extrapolations}. 
While the coloured SUSY partners 
are expected to be discovered at
the hadron colliders Tevatron and LHC, 
future $e^+e^-$ linear colliders will provide a comprehensive picture of the
weakly interacting, non--coloured particles \cite{LC,TDR}.
Moreover, the
detailed analysis of their properties will be a central target of
experiments at the linear colliders such as JLC/NLC/TESLA 
in the sub-TeV phase 
and as CLIC, 
a collider concept
for extending the energy to the multi--TeV range.

The analysis program for the new particles has been developed in great
detail for the mass parameters and mixings in the
(non--coloured) sfermion and gaugino sectors \cite{All-Parameters}. 
It has
been shown in particular how the SU(2)$\times$U(1) gaugino mass parameters
$M_2$ and $M_1$, as well as the higgsino mass parameter $\mu$ can be
extracted \cite{Choi,CKMZ}.
However, while $\tan\beta$, the mixing parameter
in the Higgs sector, can be measured well for moderate values in the
chargino/neutralino sector, only bounds can be set on $\tan\beta$ if this
parameter is large, {\it i.e.} $\tan\beta\grts 10$, for simple mathematical
reasons discussed later.

Several processes have been studied to measure $\tan\beta$ in 
different ways, complemented also by methods for measuring the
trilinear $A$ couplings in the superpotential (see {\it e.g.} 
Ref.~\cite{nachtrag} and references therein).
 They include heavy
Higgs boson decays to fermion and sfermion pairs, Higgs radiation off
fermions and sfermions, and others. 
For this purpose some of us \cite{susy02} made a detailed study of
the tau polarisation in stau production. In this report we expand on this 
work and perform
a comprehensive analysis of
polarisation effects in sfermion decays to fermions plus 
neutralinos/charginos in $e^+e^-$ pair production of third-generation 
sfermions:
\beqn
\begin{array}{llll}
e^+e^-\to \tilde{\tau}_i\bar{\tilde{\tau}}_j\, ,& &
\tilde{\tau}_i\to\tau\tilde{\chi}^0_k& \quad\mbox[i,j=1,2;\ k=1,\ldots,4]
 \, , \la{eq_stau}\\
e^+e^-\to \tilde{t}_i\bar{\tilde{t}}_j\, ,& &
\tilde{t}_i\to t \tilde{\chi}^0_k& 
\quad\mbox[i,j=1,2\ ;k=1,\ldots,4] \, ,
\la{eq_stop}\\
e^+e^-\to \tilde{b}_i\bar{\tilde{b}}_j\, ,& &
\tilde{b}_i\to t \tilde{\chi}^{-}_k& \quad\mbox[i,j=1,2;\ k=1,2] \, .
\la{eq_sbottom}
\end{array}
\eeqn
The $\tau$, $t$ fermions are longitudinally polarised in the 2--body
decays of the scalar particles -- the neutralino/chargino
spin states are not measured.

Stau production has been proposed in Ref.~\ci{Noji} 
for investigating the
properties of neutralinos. We do not only expand on this work but
rather focus on the processes (\ref{eq_stau}) 
as a means for measuring separately $\tan\beta$ and
the trilinear couplings $A_{\tau}$, $A_t$, $A_b$ in the superpotential.

These parameters enter the off--diagonal $L/R$ element of 
the sfermion mass matrix in the combination
\bequ
m_{LR}^2[\tilde{f}]=m_f \left[ A_f-\mu\tan\beta(\cot\beta) \right]
\la{eq_mixlr}
\eequ
for down (up)--type particles, respectively. 
The matrix element can be related
directly to experimental observables -- 
the physical masses $m_{\tilde{f}_1}$, 
$m_{\tilde{f}_2}$ and the mixing angle $\theta_{\tilde{f}}$:
\bequ
m_{LR}^2[\tilde{f}]=\frac{1}{2} ( m_{\tilde{f}_1}^2-m_{\tilde{f}_2}^2 ) 
         \sin 2 \theta_{\tilde{f}}
\la{eq_sinlr}
\eequ
While the sfermion masses can be determined accurately from decay
spectra and from threshold scans, the mixing angles can be extracted
from sfermion pair production.

The trilinear
$A_f$ couplings and $\tan\beta$ can be disentangled by measuring the
fermion polarisation in the decays $\tilde{f}\to f \tilde{\chi}$. In
particular for large $\tan\beta$, the properties of the charginos and
neutralinos -- masses and mixings -- are nearly independent of the 
specific value of this
parameter as the gaugino mass matrices depend solely on $\cos 2 \beta
\simeq -1+2/\tan^2\beta$ and $\sin 2 \beta \simeq 2/\tan\beta$.  
By contrast, the Yukawa
couplings $f \tilde{f} \tilde{\chi}$ for down--type particles are of
order $\cos^{-1}\beta\simeq \tan\beta$, with high sensitivity to large
$\tan\beta$ -- to the extent that the wave functions of the associated
neutralinos and charginos possess non--negligible higgsino
components.  If this is not realised for the light gauginos, sfermion
decays to heavy gauginos may be exploited in major parts of the
supersymmetry parameter space wherever those decay channels are
kinematically open and the corresponding
decay branching ratios are sufficiently large.

The polarisation of fast moving $\tau$ particles affects the shape of the
energy
spectrum in decays like $\tau^{-}\to\nu_{\tau}\pi^{-}$ by
angular--momentum conservation. For positive $\tau$ 
helicity, for instance,
the pion is emitted preferentially in forward direction, carrying a
large fraction of the $\tau$ energy 
while the spectrum is
suppressed in the opposite
direction, i.e. for soft $\pi$ energies.
Another useful analyser is provided by the
$\rho$ decay channel.
 
The polarisation of the top quark in decays $\tilde{b}\to t 
\tilde{\chi}^{-}$ and $\tilde{t}\to t \tilde{\chi}^{0}$ can be 
determined from
the distribution of the quark jets
in the hadronic top decays $t\to b \, +\,c \bar{s} $.

The report is organised as follows. In the next section we discuss the
general analysis of the $\tilde{\tau}/\tau$ sector, 
followed in the third section 
by an experimental simulation
for a specific large $\tan\beta$ reference
point, $RP$, inferred from the Snowmass Point SPS1a \cite{SPS}. 
In the fourth section we will present the analogous analysis for
$\tilde{t}$ and $\tilde{b}$ decays including details on the
measurement of the top quark polarisation in the decay final states.

\section{The {\boldmath $\tilde{\tau}/\tau$} System}
\subsection{Masses and Mixing}
Because of the large Yukawa couplings in the third generation, the
left-- and right--chiral stau states\footnote{The following paragraphs
are
included for the sake of coherence (see also Ref. \cite{Noji,stau}).}
$\tilde{\tau}_L$ and $\tilde{\tau}_R$ mix to form mass eigenstates
$\tilde{\tau}_1$ and $\tilde{\tau}_2$. The mass matrix in the L/R current 
basis can be written in the form
\begin{eqnarray}
{\cal M}^2_{\tilde{\tau}} = 
      \left( \begin{array}{cc}
        M^2_L + m_{\tau}^2  + D_L & m_{\tau}(A_{\tau} -\mu \tan\beta)  \\
 m_{\tau}(A_{\tau} -\mu \tan\beta)&  
M^2_E + m_{\tau}^2  + D_R \end{array} \right)
=\left( \begin{array}{cc}
        m_{LL}^2 & m_{LR}^2 \\
 m_{LR}^2&  m_{RR}^2 \end{array} \right)
\label{eq_21_002}
\end{eqnarray}
with the SU(2) doublet (singlet) mass parameter $M_L^2$ ($M_E^2$); the 
D--terms
$D_L = (-{\frac{1}{2}} + \sin^2 \theta_W ) \cos(2\beta) m^2_Z$ and
$D_R = - \sin^2 \theta_W \cos(2 \beta)  m^2_Z$;
the trilinear $\tilde{\tau}_R$$\tilde{\tau}_L$$H_1$ stau--Higgs 
coupling
$A_{\tau}$; the higgsino mass para\-meter $\mu$; and $\tan\beta=v_2/v_1$
the ratio
of the two Higgs vacuum expectation values. 
The mass parameters 
$m_{LL}^2$, $m_{RR}^2$ are positive for $\tan\beta>1$, whereas 
$m_{LR}^2$ may carry either sign.

The two mass eigenvalues,
\begin{equation}
m_{\tilde{\tau}_{1,2}}^2=
\frac{1}{2} \left [ m_{LL}^2+m_{RR}^2\mp
\sqrt{(m_{LL}^2-m_{RR}^2)^2+4 m_{LR}^4} \right ] , \label{eq_massev}
\end{equation}
are ordered in the sequence $m_{\tilde{\tau}_1}<m_{\tilde{\tau}_2}$
by definition.
The mixing angle 
$\theta_{\tilde{\tau}}$ rotates the current states to the mass
eigenstates,
\bequ
{\tilde{\tau}_1 \choose \tilde{\tau}_2}=
\left(\begin{array}{cc} \cos\theta_{\tilde{\tau}} & \sin\theta_{\tilde{\tau}}\\
 -\sin\theta_{\tilde{\tau}} & \cos\theta_{\tilde{\tau}} \end{array} \right)
{\tilde{\tau}_L \choose \tilde{\tau}_R}
\eequ
The stau mixing angle, defined on the
interval $0\le\theta_{\tilde{\tau}}<\pi$, is related to
the diagonal and off--diagonal elements of the mass matrix,
\begin{equation}
\tan 2 \theta_{\tilde{\tau}}= \frac{- 2 m_{LR}^2}{m^2_{RR}-m^2_{LL}} \, .
\la{eq_21_01}
\end{equation}
In reverse, the elements of the mass matrix can be expressed by the
three characteristics of the state system, i.e. the two masses and the
mixing angle:
\begin{eqnarray}
m_{LL}^2&=&\frac{1}{2} (
{m^2_{\tilde{\tau}_1}+m^2_{\tilde{\tau}_2}} )
-\frac{1}{2} (
{m^2_{\tilde{\tau}_2} - m^2_{\tilde{\tau}_1}} )
\cos 2 \theta_{\tilde{\tau}},\la{eq_m_ll}\\
m_{RR}^2&=&\frac{1}{2} (
{m^2_{\tilde{\tau}_1}+m^2_{\tilde{\tau}_2}} )
+\frac{1}{2} (
{m^2_{\tilde{\tau}_2} - m^2_{\tilde{\tau}_1}} )
\cos 2 \theta_{\tilde{\tau}},\la{eq_m_rr}\\
m_{LR}^2&=&\frac{1}{2} (m_{\tilde{\tau}_1}^2-m^2_{\tilde{\tau}_2} )
\sin 2 \theta_{\tilde{\tau}}.\la{eq_m_lr}
\end{eqnarray}
Depending on the sign of $\cos 2\theta_{\tilde{\tau}}$, the SUSY mass
parameter $m_{LL}^2$ is either smaller or larger than $m_{RR}^2$.

The two masses $m_{\tilde{\tau}_{1,2}}$ can be measured from 
the endpoints of the spectra in decay distributions and from
threshold scans in $e^+e^-$ annihilation.
The mixing angle $\theta_{\tilde{\tau}}$ can be determined from
measurements of the production cross sections 
$e^+e^-\to \tilde{\tau}_i\tilde{\tau}_j$ [i,j=1,2]:
\begin{eqnarray}
\sigma(e^+ e^-\to \tilde{\tau}_i \tilde{\tau}_j)&=& 
\frac{8 \pi \alpha^2}{3 s}
\lambda^{\frac{3}{2}}
\Big[ c_{ij}^2 \,
 \frac{|\Delta(Z)|^2}{\sin^4 2 \theta_W} \,
 ({\cal P}_{-+} L^2_{\tau}+{\cal P}_{+-} R^2_{\tau})
\nonumber\\
&&
+\delta_{ij} \frac{1}{16} ({\cal P}_{-+}+{\cal P}_{+-})
+\delta_{ij} {c_{ij}} \,
\frac{Re(\Delta(Z))}{2 \sin^2 2\theta_W} \,
 ({\cal P}_{-+} L_{\tau}+{\cal P}_{+-} R_{\tau})\Big],\label{App-59}
\end{eqnarray}
with $s$ denoting the cm energy squared. $\lambda^{\frac{1}{2}}$,
with $\lambda =
[1-(m_{\tilde{\tau}_i}+m_{\tilde{\tau}_j})^2/s]
[1-(m_{\tilde{\tau}_i}-m_{\tilde{\tau}_j})^2/s]$,
is proportional to the  
velocity of the $\tilde{\tau}$ in the final
state; the
coefficient $\lambda^{3/2}$ in the cross section is characteristic for
the $P$--wave suppression of pair-production of scalar particles at
threshold in $e^+e^-$
annihilation. $\Delta(Z)=is/(s-m^2_Z+im_Z\Gamma_Z)$ denotes the
(renormalised) $Z$
propagator. The lepton/slepton couplings include the mixing angle,
\begin{eqnarray}\label{fkop1}
&&c_{11/22}=\frac{1}{2} [L_{\tau}+R_{\tau}\pm (L_{\tau}-R_{\tau})
\cos 2\theta_{\tilde{\tau}}] \, ,    \\
&&c_{12}=c_{21}=\frac{1}{2}(L_{\tau}-R_{\tau}) 
\sin2 \theta_{\tilde{\tau}} \, ,
\label{App-65}
\end{eqnarray}
with $L_{\tau}=\left(-\frac{1}{2} +\sin^2 \theta_W \right)$ 
and $R_{\tau}=\sin^2 \theta_W$ being 
the left/right chiral $Z$ charges of $\tau$.
The electron/positron polarisation coefficients are 
defined as ${\cal P}_{-+} = (1-P_{e^-})(1+P_{e^+})$
and {\it vice versa}, with the first/second index
denoting the $e^-/e^+$ helicity, and $P_{e^-}$, $P_{e^+}$ being the 
polarisation.

The measurement of one of the diagonal cross sections, for example, 
will determine
$\cos2\theta_{\tilde{\tau}}$ up to at most a single ambiguity{\footnote{
Generally the condition $|\cos 2\theta_{\tilde{\tau}}| \leq 1$ is 
not met by both solutions of the quadratic equation (\ref{App-59})
and the analysis of the single production
channel 11 is sufficient in this case.}}.
The ambiguity
can be resolved by measuring the cross sections for two pairs
11 and 22, or
by using polarised beams, or by varying the beam energy. The second method
may be most useful in practice.
The 12  cross section is generally small and therefore less useful in 
practice.
Either of the other options will finally lead to a unique value of
$\cos 2 \theta_{\tilde{\tau}}$ from which the modulus
$|\cos\theta_{\tilde{\tau}}|$ can be derived and, equivalently,
 $\theta_{\tilde{\tau}}$
up to the reflection $\theta_{\tilde{\tau}}\leftrightarrow
\pi-\theta_{\tilde{\tau}}$. 
At the very end we are left with a sign ambiguity
in the mixing parameter $\sin 2\theta_{\tilde{\tau}}$. 

{\it In summary}. If the two masses $m_{\tilde{\tau}_{1,2}}$ 
and the mixing angle
$|\cos \theta_{\tilde{\tau}}|$
have been determined, the off-diagonal element
of the mass matrix is fixed up to a sign ambiguity.
Thus the combination
$(A_{\tau}-\mu \tan\beta)$ of the fundamental supersymmetric
parameters $A_{\tau}$ and $\tan\beta$ can be evaluated up to a simple
sign ambiguity solely  from 
measurements of masses and cross sections.

\subsection{{\boldmath ${\tau}$} Polarisation in {\boldmath
$\tilde{\tau}$} Decays}
To disentangle the parameters $A_{\tau}$ and $\tan\beta$
in the off--diagonal element of the mass matrix, the measurement
of the $\tau$ (longitudinal) polarisation in the decays
\begin{eqnarray}
&&\tilde{\tau}_{1}\to \tau \tilde{\chi}^0_k \label{eq_22_1a}
\quad\mbox{and}\quad
\tilde{\tau}_{2}\to \tau \tilde{\chi}^0_k\quad [k=1,\ldots,4]
\end{eqnarray}
proves crucial. The
$\tau$ polarisation depends in general on the mixing of the
neutralino $\tilde{\chi}^0_k$ states and it can be expressed 
in terms of the Yukawa couplings ${\mathfrak a}^{L,R}_{ik}$ \ci{Noji,susy02}:
\begin{eqnarray}
P_{\tilde{\tau}_i\to \tau \tilde{\chi}^0_k}&=&
\frac{({\mathfrak a}_{ik}^R)^2-({\mathfrak a}_{ik}^L)^2}
{({\mathfrak a}_{ik}^R)^2
+({\mathfrak a}_{ik}^L)^2} \, , \label{eq_22_2a}
\end{eqnarray}
with
\begin{eqnarray}
{\mathfrak a}_{1k}^{L,R}&=&\cos\theta_{\tilde{\tau}} {\mathfrak
a}_{Lk}^{L,R}+\sin\theta_{\tilde{\tau}} {\mathfrak a}_{Rk}^{L,R}
\quad\mbox{and}\quad
{\mathfrak a}_{2k}^{L,R}=-\sin\theta_{\tilde{\tau}} {\mathfrak a}_{Lk}^{L,R}
+\cos\theta_{\tilde{\tau}} {\mathfrak a}_{Rk}^{L,R}, 
\end{eqnarray}
where the scalar currents are defined by the interaction
\begin{equation}
{\cal L}_{\tau}=\sum_{{i=1,2 \atop k=1,\ldots,4}}
\tilde{\tau}_i( \bar{\tau}_R {\mathfrak a}^R_{ik}
+ \bar{\tau}_L {\mathfrak a}^L_{ik})\tilde{\chi}^0_k
\label{eq_22_3}
\end{equation}
in the \{gaugino; higgsino\} basis
$\{\tilde{B},\tilde{W}_3;\tilde{H}_1,\tilde{H}_2\}$:
\begin{eqnarray}
{\mathfrak a}_{Lk}^R&=&- 
\frac{g}{\sqrt{2}} \frac{m_{\tau}}{m_W \cos\beta} N_{k3}
\quad\mbox{and}\quad 
{\mathfrak a}_{Rk}^R=
- \frac{2 g}{\sqrt{2}} N_{k1} \tan\theta_W \, , \label{eq_22_4a}\\
{\mathfrak a}_{Lk}^L&=&+ \frac{g}{\sqrt{2}} [N_{k2}+N_{k1} \tan\theta_W]
\quad\mbox{and}\quad
{\mathfrak a}_{Rk}^L={\mathfrak a}_{Lk}^R\label{eq_22_4d}.
\end{eqnarray}

\begin{figure}[htb]
\setlength{\unitlength}{1cm}
\begin{center}
\begin{picture}(7,12)(4.2,0)
  \put(0,0){\includegraphics{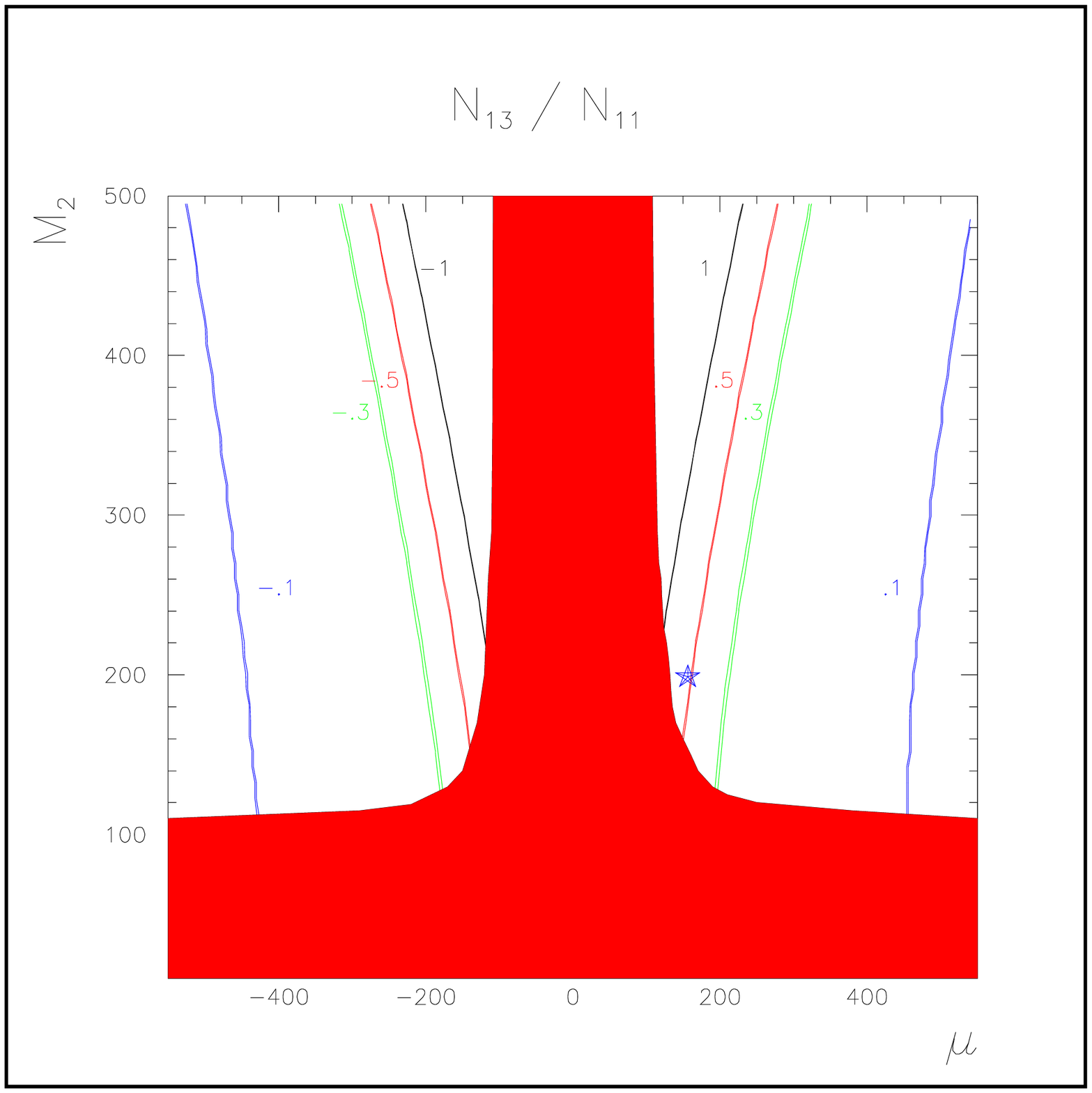}}
  \put(8,0){\includegraphics{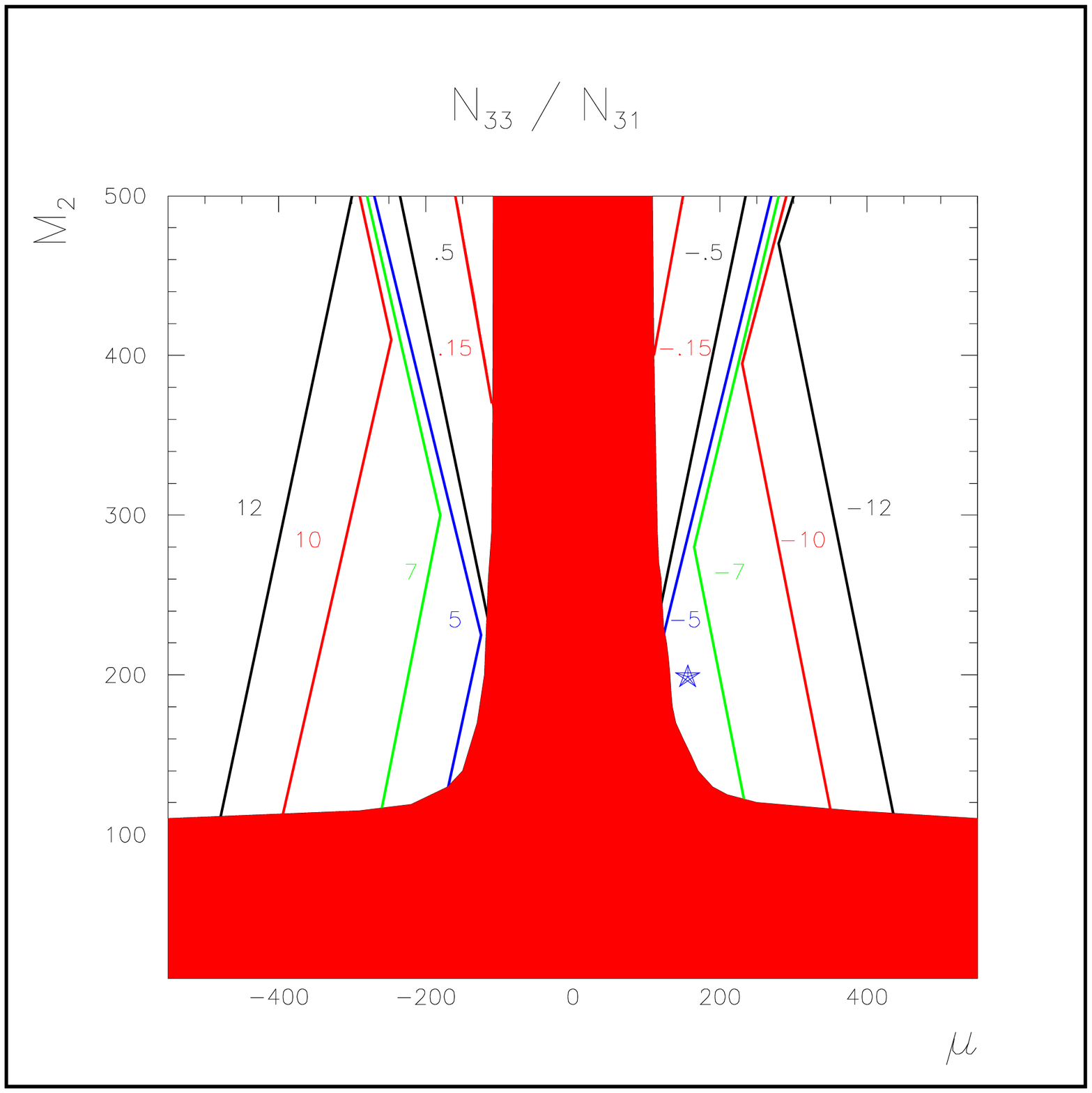}}
\end{picture}
\end{center}\vspace*{-2.5cm}
  \caption{\label{con_ratio} {\it 
      Contour plots in the $(M_2, \mu)$ plane
      for the matrix elements $N_{13}/N_{11}$ and $N_{33}/N_{31}$. The 
      reference point $RP$ is indicated by the blue star.
      The exclusion bounds are set to
      $m_{\tilde{\chi}^0_1} \ge 45$~GeV and 
      $m_{\tilde{\chi}^{\pm}_1} \ge 103$~GeV.}}%
\end{figure}

The elements of the neutralino mixing matrix $N_{km}$ approach
asymptotically a value independent of high $\tan\beta$ so that
the coefficients in front of the 
${\mathfrak a}^{R}_{Lk}$ and ${\mathfrak a}^{L}_{Rk}$ couplings
are the key elements for our purpose. They are
proportional to $\cos^{-1}\beta\simeq \tan\beta$ with a coefficient that
depends only on parameters measured in the chargino/neutralino
sector but is nearly independent of $\tan\beta$. 
To exploit the strong $\tan\beta$
dependence of ${\mathfrak a}^{R}_{Lk}$ and 
${\mathfrak a}^{L}_{Rk}$, a significant higgsino
component $\sim N_{k3}$ must be present in the neutralino wave function.  
Therefore $\tilde{\tau}_i$ [i=1,2] may be needed both to cover also 
the heavy neutralino decay channels. 
Unitarity ensures
that at least one neutralino state with significant higgsino component
in the wave function will be accessed. Since 
the coefficients of the key elements 
depend only on already measured parameters, we can  
predict {\it a priori} to what extent the method is useful for a 
particular decay process.
The contour plots in Fig.~\ref{con_ratio} exemplify typical
values of the higgsino parameter $N_{13}/N_{11}$  of $\tilde{\chi}^0_1$
{\it versus} $N_{33}/N_{31}$ of $\tilde{\chi}^0_3$ (both normalised 
to the bino component) in the $(\mu,M_2)$ plane of the MSSM.
[The constraint relation
$M_1/M_2=\frac{5}{3} \tan\theta_W^2$ is adopted as an
auxiliary assumption just for the sake of 
simplicity.]
The kinks of the contour curves $N_{33}/N_{31}$ 
in Fig.~\ref{con_ratio}b are caused by the exchange of the 
gaugino/higgsino mixing character of
$\tilde{\chi}^0_2$ and $\tilde{\chi}^0_3$ as the 
corresponding mass eigenvalues change their ordering, cf. 
\cite{Bartl:1989ms}.
 The reference point $RP$, given in Table~\ref{tab_ref}, is 
marked by a star.

Separating the neutralino mixing parameters from the relevant Yukawa 
couplings,
\begin{eqnarray}
n_g&=&1+\cot\theta_W\frac{N_{12}}{N_{11}} \, ,
\label{eq_22_6b}\\
n_h&=&\cot\theta_W\frac{N_{13}}{N_{11}} \, , \label{eq_22_6e}
\end{eqnarray}
gives rise to a transparent representation for the $\tau$ polarisation in 
$\tilde{\tau}_1\to \tau\tilde{\chi}^0_1$ decay:
\begin{eqnarray}
P_{\tilde{\tau}_1\to \tau\nt_1}&=&\frac{(4-n_g^2)-(4+n_g^2-2
n_h^2 \mu_{\tau}^2/\cos^{2}\beta)
\cos{2\theta_{\tilde{\tau}}}+2(2+n_g) n_h \sin{2\theta_{\tilde{\tau}}}
 \mu_{\tau}/\cos\beta}
{(4+n_g^2+2 n_h^2 \mu_{\tau}^2/\cos^{2}\beta)
-(4-n_g^2)
\cos{2\theta_{\tilde{\tau}}}+2(2-n_g) n_h \sin{2\theta_{\tilde{\tau}}}
\mu_{\tau}/\cos\beta}
\label{eq_22_5a}
\end{eqnarray}
with the abbreviation $\mu_{\tau}=m_{\tau}/m_W$.
The formula is easily transcribed to other neutralino $\tilde{\chi}^0_k$
decays by adjusting the mixing coefficients $N_{1m} \Rightarrow N_{km}$.
The transition to
$\tilde{\tau}_2$ just requires flipping the signs of
$\sin 2\theta_{\tilde{\tau}}$ and $\cos 2\theta_{\tilde{\tau}}$.
Note that the polarisation itself is independent of the stau and
neutralino masses.

\section{A Specific Example}

To study the experimental feasibility of measuring the parameters $\tan\beta$
and $A_{\tau}$,  
we have defined the
reference point $RP$ in Table~\ref{tab_ref}, that is 
motivated\footnote{We have checked, by adopting the programme 
\cite{Microomegas}, that the point $RP$, whose parameters are given in 
Table~\ref{tab_ref}, is in agreement with constraints from
present data for $[g-2]_{\mu}$, $b\to s \gamma$ 
and $\Omega$.}
by the Snowmass point SPS1a \cite{SPS}. 
The particle masses associated with the reference 
point $RP$ are collected in
Table~\ref{tab_mass}. The matrices diagonalizing the 
neutralino and chargino mass
matrices are displayed in
Tables~\ref{tab_evneut} and \ref{tab_evchar}.  
Note that the lightest neutralino
$\tilde{\chi}^0_1$ state (and also the 
chargino $\tilde{\chi}^{+}_1$) has a significant
higgsino component, $N_{13}$ (and $V_{12}$).
In Table~\ref{tab_rates} the cross sections 
$\sigma(\ee\to\tilde{\tau}_{i}\tilde{\tau}_{j})$  are listed 
for $\sqrt{s}=500$~GeV and $800$~GeV 
and various $e^{\pm}$ beam polarisations  $P_{e^-}$ and $P_{e^+}$.
The predicted $\tau$ polarisations and branching ratios of the 
decays $\stau_{1,\,2}\to \tau\nt_k$ are collected in 
Table~\ref{tab_poltheo}. 

For moderate values of $\tan\beta$, the polarisation is affected only
indirectly by the wave-function through the gaugino mixing $n_g$ while
the direct dependence on $\tan\beta$ through the Yukawa coupling is
suppressed by the small mass ratio $\mu_{\tau}\sim 10^{-2}$.  By
contrast for `large $\tan\beta$' in the range from 10 to 50, the
gaugino and higgsino parameters, $n_g$ and $n_h$, are nearly
independent of $\tan\beta$, cf. Ref. \cite{CKMZ} and Appendix A,
and the mixing angle $\theta_{\tilde{\tau}}$, with 
$\tan 2\theta_{\tilde{\tau}} \sim 2 m_{\tau}\mu\tan\beta/M_{E,L}^2$,
is expected to be still sufficiently away from the
asymptotic value $\pi/4$. In this range, the polarisation is affected
strongly by the Yukawa coupling $\sim \/{\cos^{-1}\beta} \simeq \tan\beta$ so
that the polarisation measurement provides us with a new experimental 
method for determining large values of $\tan\beta$.

The Yukawa couplings can be proven as the origin for the sensitivity 
of the polarisation on $\tan\beta$ 
in $\tilde{\tau}_1\to \tau\tilde{\chi}^0_1$. This can be demonstrated by
comparing the exact value of the $\tau$ polarisation $P=0.85$ with 
the approximate value derived for
the gaugino and higgsino components $n_g$, $n_h$ of the neutralino 
in the asymptotic limit $\tan\beta\to\infty$: $P_{\infty}=0.86$.  
\begin{table}[p]
\begin{minipage}{8cm}
\begin{center}
\begin{tabular}{|l|c|c|}
\hline
\multicolumn{3}{|c|}{Basic $RP$ Parameters}\\
\hline\hline
Gaugino Masses 
& $M_1$ &\ 99.1~GeV\\
& $M_2$ & 192.7~GeV\\ \hline
Higgs(ino) Parameters 
& $\mu$ & 140~GeV\\
& $\tan\beta$& 20 \\ \hline
Slepton Mass Parameters 
& $M_L$ & 300~GeV \\
& $M_E$ & 150~GeV\\
Squark Mass Parameters 
& $M_Q$ & 596~GeV\\ 
$3^{rd}$ generation
& $M_U$ & 525~GeV\\   
& $M_D$ & 617~GeV \\ \hline 
Trilinear Couplings 
& $A_{\tau}$ & $-254$~GeV  \\
& $A_b$      & $-773$~GeV\\
& $A_t$      & $-510$~GeV\\ \hline
\end{tabular}
\caption{{\it 
    Definition of the reference scenario $RP$ 
    at the electroweak scale} 
  \label{tab_ref}}
\end{center}
\end{minipage}\hspace{1.5cm}
\begin{minipage}{7cm}
\begin{center}
\begin{tabular}{|l|c|c|}
\hline
\multicolumn{3}{|c|}{Masses and Sfermion Mixing Angles}\\ \hline\hline
Neutralinos 
& $m_{\tilde{\chi}^0_1}$ &\ 78~GeV\\
& $m_{\tilde{\chi}^0_2}$ & 126~GeV\\
& $m_{\tilde{\chi}^0_3}$ & 152~GeV\\
& $m_{\tilde{\chi}^0_4}$ & 240~GeV\\ \hline
Charginos 
& $m_{\tilde{\chi}^{\pm}_1}$ & 110~GeV\\
& $m_{\tilde{\chi}^{\pm}_2}$ & 240~GeV\\ \hline
Staus 
& $m_{\tilde{\tau}_1}$ & 155~GeV\\
& $m_{\tilde{\tau}_2}$ & 305~GeV\\ \hline
Sbottoms 
& $m_{\tilde{b}_1}$ & 592~GeV\\
& $m_{\tilde{b}_2}$ & 624~GeV\\ \hline
Stops 
& $m_{\tilde{t}_1}$ & 497~GeV\\
& $m_{\tilde{t}_2}$ & 665~GeV\\ \hline \hline
Sfermion Mixing Angles
& $\theta_{\stau}$    & 1.492 \\
& $\theta_{\tilde b}$ & 0.485 \\
& $\theta_{\tilde t}$ & 0.987 \\ \hline
\end{tabular}
\caption{{\it 
    Physical masses and sfermion mixing angles in the reference 
    scenario $RP$ \label{tab_mass}}}
\end{center}
\end{minipage}
\end{table}

\begin{table}[p]
\begin{minipage}{8cm}
\begin{center}
\begin{tabular}{|c|cccc|}
\hline
\multicolumn{5}{|c|}{Neutralino Mixing Matrix}\\   \hline\hline
&$N_{k1}$&$N_{k2}$ &$N_{k3}$ &$N_{k4}$ \\ \hline
$\tilde{\chi}^0_1$ & $-0.730$    & $\ \ 0.248$ & $-0.548$    & $\ \ 0.325$\\
$\tilde{\chi}^0_2$ & $\ \ 0.657$ & $\ \ 0.488$ & $-0.425$    & $\ \ 0.386$\\
$\tilde{\chi}^0_3$ & $\ \ 0.118$ & $-0.161$    & $-0.660$    & $-0.724$\\
$\tilde{\chi}^0_4$ & $-0.147$    & $\ \ 0.821$ & $\ \ 0.289$ & $-0.470$ 
\\ \hline
\end{tabular}
\caption{{\it 
    Neutralino mixing matrix of $RP$} \label{tab_evneut}}
\end{center}
\end{minipage} \hspace{1.cm}
\begin{minipage}{8cm}
\begin{center}
\begin{tabular}{|c|cc||c|cc|}
\hline
\multicolumn{6}{|c|}{Chargino Mixing Matrix}\\  \hline \hline
& $V_{k1}$ & $V_{k2}$ & & $U_{k1}$ & $U_{k2}$\\\hline
$\tilde{\chi}^{+}_{1}$ &\  0.669 & $-0.743$ & 
$\tilde{\chi}^-_1$     & $-0.408$ &\ 0.913\\
$\tilde{\chi}^{+}_2$   &\  0.743 &\ \ $0.669$ 
& $\tilde{\chi}^-_2$   &\ \ $0.913$ &\ 0.408\\ \hline
\end{tabular}
\caption{{\it 
    Chargino mixing matrix of $RP$} \label{tab_evchar}}
\end{center}
\end{minipage}
\end{table}

\begin{table}[p]
\begin{center}
\begin{tabular}{|c||c||c|c|c|}
\hline
             & $\sqrt{s}=500$~GeV \  & 
\multicolumn{3}{|c|}{$\sqrt{s}=800$~GeV}\\ \hline\hline
$(P_{e^-},P_{e^+})$ & $\sigma(e^-e^+\to\tilde{\tau}_1\tilde{\tau}_1)$ & 
$\sigma(e^-e^+\to\tilde{\tau}_1\tilde{\tau}_1)$ & 
$\sigma(e^-e^+\to\tilde{\tau}_2\tilde{\tau}_2)$ & 
$\sigma(e^-e^+\to\tilde{\tau}_1\tilde{\tau}_2)$\\ \hline
unpolarised   &46.8 fb  &29.4 fb & 12.4 fb & 0.06 fb\\
$(-0.8,0)$    &24.0 fb  &15.3 fb & 19.1 fb & 0.07 fb \\
$(+0.8,0)$    &70.0 fb  &43.6 fb &  5.7 fb & 0.05 fb \\
$(-0.8,+0.6)$ & 29.3 fb &18.9 fb & 30.0 fb & 0.11 fb \\
$(+0.8,-0.6)$ &109.1 fb &68.3 fb &  6.7 fb & 0.07 fb\\ \hline
\end{tabular}
\caption{{\it 
    Production cross sections of $e^+e^-\to\stau_i\stau_j$ 
    for reference point $RP$ with polarised beams
    $(P_{e^-},\,P_{e^+})$.
    The cross sections for
    $\tilde{\tau}_1\tilde{\tau}_2$ production 
    at $\sqrt{s}=500$~GeV are less than $0.1$~fb} 
\label{tab_rates}}
\end{center}
\end{table}

\begin{table}[htb]
\begin{center}
\begin{tabular}{|c|c|c||c|c|}
\hline
\multicolumn{5}{|c|}{$\tau$ Polarisations and $\stau$ Branching Ratios}
\\ \hline\hline
& \multicolumn{2}{| c||}{$\tilde{\tau}_1\to \tau \,\tilde{\chi}^0_k$} &
\multicolumn{2}{|c|}{$\tilde{\tau}_2\to \tau \,\tilde{\chi}^0_k$}\\
\hline
   & $P_\tau$ & $\cB_{\stau_1\to\tau\nt}$ 
   & $P_\tau$ & $\cB_{\stau_2\to\tau\nt}$
\\ \hline
$\tilde{\chi}^0_1$ & $+0.85$ & $0.78$ & $+0.11$ & $0.04$ \\
$\tilde{\chi}^0_2$ & $+0.76$ & $0.12$ & $-0.84$ & $0.43$ \\
$\tilde{\chi}^0_3$ &   ---   &  ---   & $+0.72$ & $0.05$ \\
$\tilde{\chi}^0_4$ &   ---   &  ---   & $-0.93$ & $0.07$ \\ \hline
\end{tabular}
\end{center}
\caption{{\it 
    $\tau$ polarisations $P_\tau$ and branching ratios $\cB_\stau$ 
    of the decays
    $\tilde{\tau}_{1,2} \to \tau \tilde{\chi}^0_k$ 
  for reference point~$RP$ \label{tab_poltheo}}}
\end{table}

\subsection{Summary of {\boldmath $\stau$} Parameter Measurements}

The parameters of the $\stau$ system can be determined by measurements
of the $\stau_1,\ \stau_2$ masses and the mixing angle $\theta_\stau$ 
from which $\tan\beta$ and the trilinear coupling $A_\tau$ can be
derived. Maximum sensitivity is achieved by proper choices of collision 
energy and beam polarisations, see Table~\ref{tab_rates}.
The following configurations with large rates are considered:
\begin{eqnarray}
  e^+_L e^-_R & \to & \stau_1 \stau_1 \  
  \qquad {\rm at} \ \sqrt{s}=500~\GeV \ ,
  \label{stau11} \\
  e^+_R e^-_L & \to & \stau_2 \stau_2  \
  \qquad  {\rm at} \ \sqrt{s}=800~\GeV \ .
  \label{stau22} 
\end{eqnarray}

Methods to determine particle masses include measurements of decay spectra  
and cross sections at the production thresholds. 
A Monte Carlo simulation of reaction~(\ref{stau11}),
described in Appendix~\ref{mcstudy}, shows that the $\stau_1$ mass can
be measured with an accuracy of $\delta m_{\stau_1}=0.5~\GeV$.
Applying extrapolations of the present and previous studies~\cite{TDR} 
to $\stau_2\stau_2$ production at higher energies,
one expects an uncertainty of $\delta m_{\stau_2}\simeq 2-3~\GeV$ 
on the $\stau_2$ mass.

\begin{figure} 
\begin{center}
\setlength{\unitlength}{1cm}
\begin{picture}(7,8)(0,-0.5)
\put(5.8,-.2){\Large$\sigma(\tilde{\tau}_1\tilde{\tau}_1)$/fb}
\put(-3,6.5){\Large$\cos2\,\theta_{\tilde{\tau}}$}
\put(-2,0){\epsfig{file=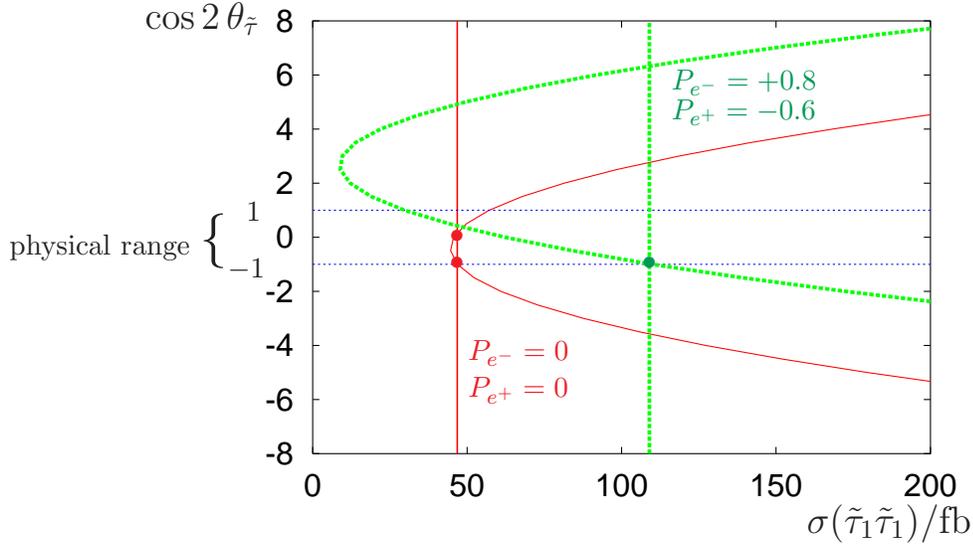,scale=.8}}
\put(1.3,2.1){\color{Red}$P_{e^-}=0$}
\put(1.3,1.6){\color{Red}$P_{e^+}=0$}
\put(1.05,3.65){\color{Red}$\bullet$}
\put(1.05,3.3){\color{Red}$\bullet$}
\put(4,5.7){\color{Green}$P_{e^-}=+0.8$}
\put(4.,5.3){\color{Green}$P_{e^+}=-0.6$}
\put(3.61,3.3){\color{Green} $\bullet$}
\put(-4.8,3.5){physical range {\huge $\{$}}
\put(-1.65,3.95){1}
\put(-1.9,3.2){$-1$}
\end{picture} 
  \caption{\label{sigpol} {\it
      Mixing angle $\cos 2\theta_\stau$ versus cross section 
      $\sigma(e^+ e^- \to \stau_1\stau_1)$ at $\sqrt{s}=500~\GeV$ 
      for beam polarisations $P_{e^-}=+0.8$ and  $P_{e^+}=-0.6$ (green) 
      and the unpolarised case (red) in the scenario $RP$.
      The vertical lines indicate the predicted cross sections.
      For unpolarised beams one observes a 
      two-fold ambiguity in $\cos2\theta_{\tilde{\tau}}$ (red dots); 
      for polarised beams, 
      however, only one solution lies in the allowed range (green dot).}}
\end{center}
\end{figure}

The $\stau$ mixing angle $\theta_\stau$ is related to the total cross
section, given by eqs.~(\ref{App-59}) -- (\ref{App-65}), and it is
displayed in Fig.~\ref{sigpol} for $\stau_1\stau_1$ production.  The
measurement of the cross section will not be limited by statistics, but
rather by systematic effects with a typical error of
$\delta\sigma/\sigma = 3\,\% $.  Using the theoretical prediction
of $\sigma(e^+_L e^-_R\to\stau_1\stau_1)= 109~$fb at the Born level for
$P_{e^-}=+80\%$, $P_{e^+}=-60\%$, with a statistical error of
1.0~fb and a systematical error of 3.3~fb, the mixing angle can be
estimated to an accuracy of $\cos 2 \theta_\stau = -0.987 \pm 0.02 \pm
0.06$ for an integrated luminosity of $\cL = 500~\fbi$, see
Fig.~\ref{sigpol}.  (The detailed simulation will be presented in
Appendix~\ref{mcstudy}.)
%

\begin{figure}
\setlength{\unitlength}{1cm}
\begin{center}
\begin{picture}(7,9)
\put(-.2,0){\includegraphics{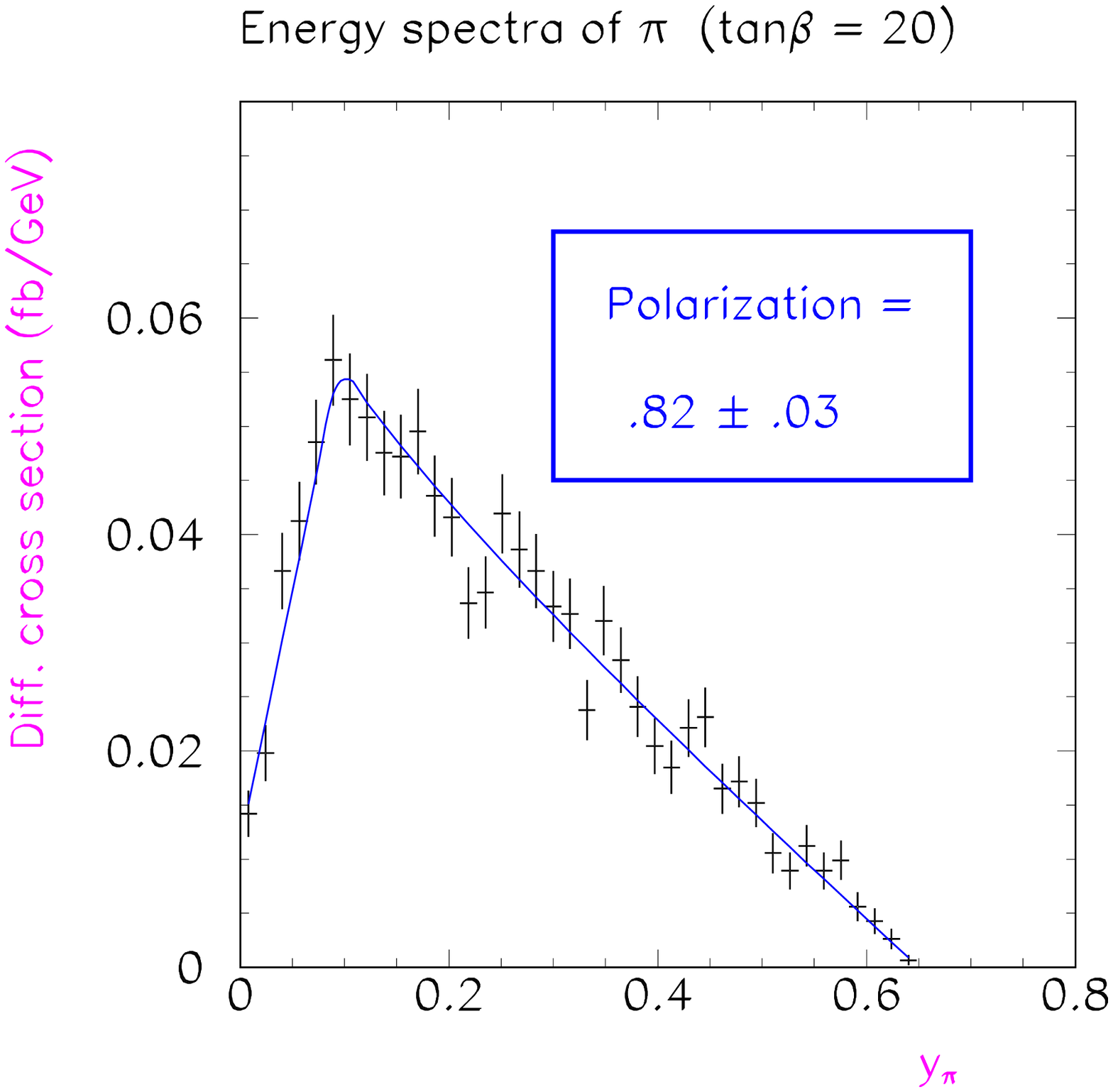}}
\put(2.,6){\epsfig{file=box.eps,width=60mm,height=40mm}}
\put(3.,1){\epsfig{file=box.eps,width=100mm,height=10mm}}
\put(-2.,9.5){\epsfig{file=box.eps,width=150mm,height=10mm}}
\put(-5.,2){\epsfig{file=box.eps,width=20mm,height=100mm}}
\put(6.6,1.2){{\Large $y_\pi=2 E_\pi/\sqrt{s}$}}
\put(-5.2,9.){{\Large
      {$\mathrm d\sigma/ \mathrm d y_{\pi} \ [\fb / \GeV]$}
}}
\end{picture}\par\vspace{0.5cm}\hspace{-1cm}
\end{center}\vspace{-2.5cm}
  \caption{\label{pi_polarization}{\it
    Pion energy spectrum $y_\pi = 2\,E_\pi/\sqrt{s}$ from 
    $\tau\to\pi\nu$ decays of 
    $e^+_L e^-_R \to \stau^+_1\stau^-_1 
    \to\tau^+\nt_1 + \tau^-\nt_1$ production with $P_{e^-}=+0.8$, 
    $P_{e^+}=-0.6$ at $\sqrt{s}=500~\GeV$,
    corresponding to $\cL=500~\fbi$;
    reference scenario $RP$.
    The curve represents a fit to a $\tau$ polarisation of
    $P_\tau = 0.82\pm0.03$. }}
\end{figure}

\begin{figure}
\setlength{\unitlength}{1cm}
\begin{center}
\begin{picture}(7,9)
\put(-2.7,8.3){\Large$\tan\beta$}
\put(9.3,1.8){\Large$P_{\tilde{\tau}_1\to\tau\nt_1}$}
\put(-.2,0){\includegraphics{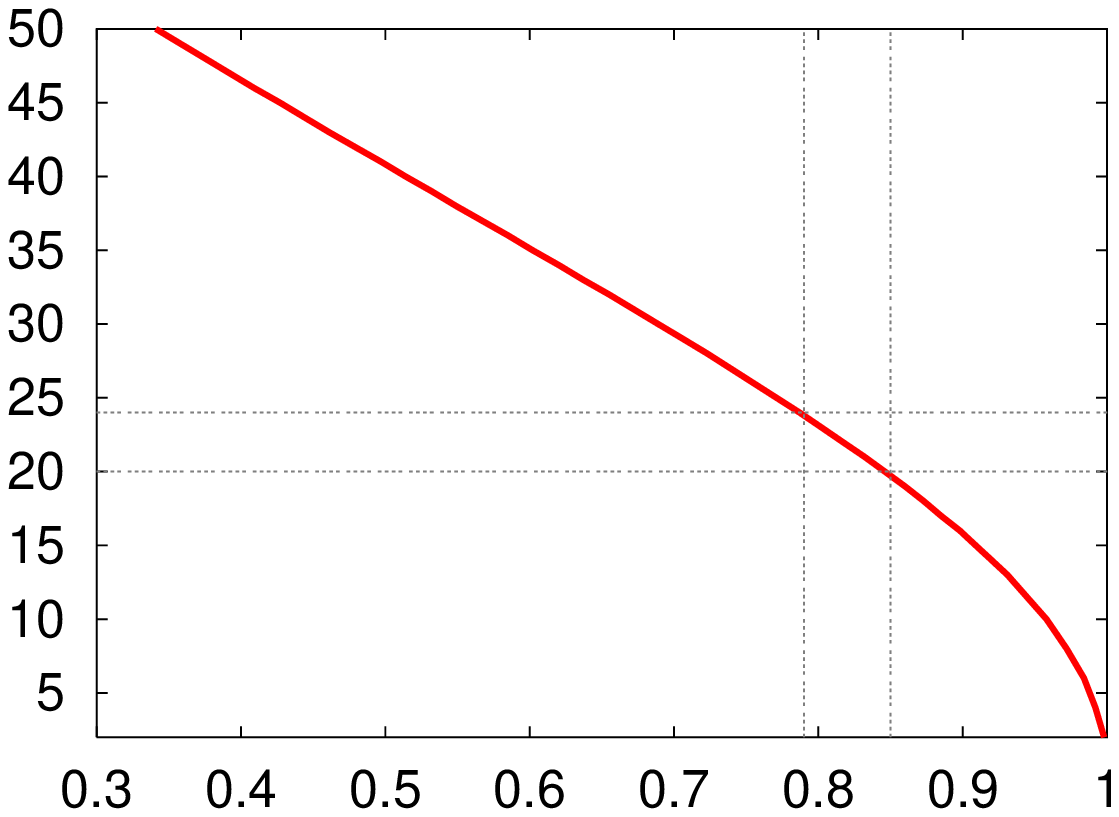}}
\end{picture}\par\vspace{.4cm}\hspace{-1cm}
\end{center}\vspace{-3cm}
  \caption{\label{tau_pol}{\it
      $\tan\beta$ versus $\tau$ polarisation $P_{\stau_1\to\tau\nt_1}$
      for the reference scenario $RP$. 
      The bands illustrate a measurement of 
      $P_{\tau}=0.82\pm0.03$ leading to 
      $\tan\beta=22\pm 2$.}}%
\end{figure}

The $\tau$ polarisation $P_\tau$ in $\stau$ decays
can be measured from the shape of the energy 
spectra of hadronic decays, e.g. $\tau\to\pi\nu$.
The energy distributions including
the complete spin correlations for the final state
$e^+_Le^-_R\to\stau_1\stau_1 \to \stau_1^\pm + \pi^\mp \nu\nt_1$
have been calculated using {\sc CompHEP}~\cite{comphep}.
The polarised $\tau$ decays have been checked to agree with 
{\sc Tauola}~\cite{TAUOLA}.
In a case study
the accuracy of a polarisation measurement has been estimated by
generating unweighted events at $\sqrt{s}=500~\GeV$
corresponding to an integrated luminosity of $\cL=500~\fbi$.
Taking all branching ratios into account and assuming an efficiency
of $\varepsilon\simeq 0.30$ 
gives rise to about 3,300 decays $\tau\to\pi\nu$ with 
the pion energy spectrum shown in Fig.~\ref{pi_polarization}.
The scaled pion energy distribution, 
$y_{\pi}=2 E_{\pi}/\sqrt{s}$, is given \cite{Noji} by  
\begin{eqnarray}
\frac{1}{\sigma}\frac{d\sigma}{dy_{\pi}} & = & 
 \frac{1}{x_{+}-x_{-}} \label{eq_fit}
\begin{cases}
  (1-P_{\tau})\, \log\frac{x_{+}}{x_{-}} +
  2P_{\tau}\, y_{\pi}\,(\frac{1}{x_{-}}-\frac{1}{x_{+}})
  &\text{$0<y_{\pi}<x_{-}$}\\[1.ex]
  (1-P_{\tau})\, \log\frac{x_{+}}{y_{\pi}} +
  2P_{\tau}\, (1-\frac{y_{\pi}}{x_{+}})
  &\text{$x_{-}<y_{\pi}<x_{+}$} 
\end{cases}  \\[1ex]
\mbox{where} \nonumber \\
  x_{+ / -} & = & \frac{m_\stau}{\sqrt{s}}
                  \left ( 1 - \frac{m_{\cx}^2}{m_{\stau}^2} \right ) 
                  \frac{1 \pm \beta}{\sqrt{1-\beta^2} } 
   \  \mbox{ with } \
  \beta \ = \ \sqrt{1-4m_{\stau}^2/s} \ . \nonumber
\end{eqnarray}  
A fit of the analytical formula to the generated spectrum
gives a polarisation of
$P_{\tilde{\tau}_1\to \tau \tilde{\chi}^0_1}= 0.82 \pm 0.03$,
compatible with the theoretical value $P_\tau^{th}=0.85$ 
quoted in Table~\ref{tab_poltheo}.
From such a measurement
of $P_{\tilde{\tau}_1\to \tau \tilde{\chi}^0_1}$ the
inversion of expression~(\ref{eq_22_5a}) leads to a 
determination of $\tan\beta = 22 \pm 2$, 
as illustrated in Fig.~\ref{tau_pol}.
Note that the above approach neglects detector acceptances
and resolution effects as well as backgrounds.
It nevertheless provides a valuable estimate of the  
precision which can be achieved and which is later supported 
by detailed simulations described in Appendix~\ref{mcstudy}.

The off-diagonal elements of the mass matrix,
eqs.~(\ref{eq_21_002}) and (\ref{eq_m_lr}),
offer, in principle, the possibility to derive the trilinear coupling
$A_{\tau}$ from the data:
\begin{equation}
  A_{\tau}=\frac{m_{\tilde{\tau}_1}^2-m_{\tilde{\tau}_2}^2}{2 m_{\tau}}
         \, \sin 2\theta_{\tilde{\tau}}
         + \mu \tan\beta \ .
  \label{atau} 
\end{equation}
In the reference scenario $RP$, with 
the trilinear coupling $A_{\tau}= -254~\GeV$, 
this method cannot be applied, however.
The second term contributes to the total error with
$\delta A_{\tau}^{(2)} = 280~\GeV$.
But the first term gives a huge error of $\delta A_{\tau}^{(2)} = 2,400~\GeV$,
even if only the statistical uncertainty in 
$\sin 2\theta_{\stau}= 0.158\pm0.125$ is taken into account.
This is an artifact of 
the large $\stau$ mass splitting compared to $m_\tau$ and
the small mixing, $\theta_{\stau}\simeq \pi/2$ in $RP$.
The situation improves considerably for models with small mass differences 
and larger mixing. 
For example, a reduction of the slepton mass parameter 
$M_L=200~\GeV \,(\simeq m_{\stau_2})$, while the other $RP$ 
parameters are left unchanged, in particular $m_{\stau_1}=155~\GeV$
(cf. Table~\ref{tab_ref}) leads to
$\sin 2\theta_{\stau}= 0.517 \pm 0.033$ and a corresponding error of
$\delta A_{\tau}^{(1)} = 200~\GeV$, comparable to the contribution of
$\delta A_{\tau}^{(2)}$ in eq.~(\ref{atau}).

\section{The {\boldmath$\tilde{t}$, $\tilde{b}\to t$} System}

The analyses presented in the previous section can easily be expanded
to squarks/quarks of the third generation. Three
points should be noted for the transition from the lepton to the quark
sector:
i) Since squarks are significantly heavier than staus, the
decays to heavier neutralino/chargino states
$\tilde{\chi}^0_{3,4}$ and $\tilde{\chi}^{\pm}_2$ are possible which,
in particular in mSUGRA--like scenarios, 
may carry a dominant higgsino component.
ii) Decays of $b$ quarks cannot be exploited, as depolarisation
effects during the fragmentation process $b\to B, \; B^{*}$ wash out
the $b$ helicity signal.  
iii) Hadronic top decays can efficiently be
used as analysers for the top polarisation in the decay 
$t\to b W\to b + c \bar{s}$ by tagging the $b$ and $c$ quarks while the
flavour of the final $\bar{s}$ jet, 
which corresponds to the charged lepton in the leptonic 
$W$ decays of Ref.~\cite{boos2}, need not be identified.  
These arguments lead us naturally to consider the channels
\beqn
\tilde{t}_i &\to& t\,\tilde{\chi}^0_k\qquad 
[i=1,2; \ k=1,\ldots,4] \ , \label{eq_stpdec}\\
\tilde{b}_i &\to& t\,\tilde{\chi}^{-}_k\qquad [i=1,2; \ k=1,2] \ .
\label{eq_stbdec}
\eeqn

From the off--diagonal elements of the $\tilde{t}$ and $\tilde{b}$
mass matrices
\beqn
  m_{LR}^2[\tilde{t}] & = &
  \frac{1}{2} (m_{\tilde{t}_1}^2-m_{\tilde{t}_2}^2)
    \sin 2 \theta_{\tilde{t}} \ = \
  m_t (A_t-\mu\cot\beta) \ , \label{offdiag_t}\\
  m_{LR}^2[\tilde{b}] & = &
  \frac{1}{2} (m_{\tilde{b}_1}^2-m_{\tilde{b}_2}^2)
    \sin 2 \theta_{\tilde{b}} \ = \
  m_b (A_b-\mu\tan\beta) \  \label{offdiag_b}
\eeqn
combinations of $\tan\beta$ and the trilinear couplings $A_t$ and $A_b$ 
can be determined.  
The sensitivity to $\tan\beta$ in the stop sector is low for
large $\tan\beta$, so that access is provided primarily to $A_t$. 
Conversely, the pattern in the sbottom sector is quite analogous to
the stau sector.

With the couplings defined in analogy to eq.~(\ref{eq_22_3})
\beqn
{\cal L}_q& = &\sum_{{i,k}}
\tilde{q}_i (\bar{t}_R {\mathfrak b}^R_{ik}+\bar{t}_L {\mathfrak b}^L_{ik})
\tilde{\chi}_k,
\label{appd_22_3}
\eeqn
the (longitudinal) polarisation formulae are modified 
slightly owing to the large top mass in the final state:
\bequ
P_{\tilde{q}_i\to t {\tilde{\chi}}_k}
=\frac{[({\mathfrak b}_{ik}^R)^2-({\mathfrak b}_{ik}^L)^2] \,f_1}
{[({\mathfrak b}_{ik}^R)^2+({\mathfrak b}_{ik}^L)^2]-2{ \mathfrak b}_{ik}^R
{\mathfrak b}_{ik}^L\,f_2}.
\label{pol_gen}
\eequ
The additional coefficients $f_1$ and $f_2$, cf. eq.~(\ref{eq_22_2a}), 
purely kinematical in origin, are given by
\bequ
f_1\ =\ m_t\, \frac{(p_{\tilde{\chi}}\, s_t)}{(p_t\,
  p_{\tilde{\chi}})} \ ,
\qquad
f_2\ =\ m_t\, \frac{m_{\tilde{\chi}}} {(p_t\, p_{\tilde{\chi}})} \ ,
\label{eq_cov}
\eequ
where $m_t$, $p_t$, $s_t$ denote the mass, momentum and longitudinal
spin vector of the decaying top quark, and $p_{\tilde{\chi}}$ the momentum of
the neutralino.
These coefficients can be written in the $t$ rest frame as
\begin{eqnarray}
f_1 & = & 
\frac{
  \lambda^{\frac{1}{2}}(m_{\tilde{q}}^2,m_{t}^2,m_{\tilde{\chi}}^2)}
     {m_{\tilde{q}}^2-m_{t}^2-m_{\tilde{\chi}}^2}
\label{eq_f1} \ ,\qquad
f_2 \ = \ \frac{2 m_{t} m_{\tilde{\chi}}}
  {m_{\tilde{q}}^2-m_{t}^2-m_{\tilde{\chi}}^2} \ .
\label{eq_f2}
\end{eqnarray} 
They approach $f_1\to 1$ and $f_2\to 0$ for small decay fermion   
masses, leading back to eq.~(\ref{eq_22_5a}).

The differential distribution of the $\bar{s}$ jet in the top 
decay is given by
\begin{eqnarray}
  \frac{1}{\Gamma}
  \frac{d \Gamma}{d\cos\theta^*_s} & = &
  \frac{1}{2} (1+P_{\tilde{q}_i\to t \tilde{\chi}_k}
  \cos\theta^*_s) ,  \label{tdecay}
\end{eqnarray}
where $\theta^*_s$ is the angle between
the $\bar{s}$ quark from the $W$-boson in the $t$ decay 
and the primary sfermion $\tilde{t}_i$ ($\tilde{b}_i$)
in the top rest frame.
\\[.5em] 

\noindent 
{\bf i) {\boldmath $\tilde{t}\to t$} Transition}\\[1ex]
The polarisation of the $t$ quark in this decay process is explicitly 
given by
\begin{eqnarray}
P_{\tilde{t}_1\to t \tilde{\chi}^0_k}&=&\frac{{\cal F}^{N} f_1}
{{\cal F}^{D_1}-{\cal F}^{D_2}  f_2} \ ,
\label{appd_5a}
\eeqn
where the coefficients of the numerator ${\cal F}^N$ and denominator 
${\cal F}^{D_1}$
are known from the massless case, only with  
charge $e_t=2/3$, Yukawa coupling $Y_t=\mu_t/(\sqrt{2} \sin\beta)$ 
and top--type electro-weak isospin $I_{3L}^{t}=1/2$ adapted properly: 
\beqn
{\cal F}^{N}&=&{I_{3L}^t}^2({n_g}-2)^2+2 e_t I_{3L}^t(n_g-2)
+\sqrt{2} \sin{2\theta_{\tilde{t}}} Y_t{n_h^t} [2 e_t+I_{3L}^t (n_g -2)]
\nonumber\\
&&+\cos{2\theta_{\tilde{t}}}[
{I_{3L}^t}^2(n_g-2)^2+2 e_t I_{3L}^t(n_g-2)+2 e_t^2-Y_t^2{n_h^t}^2]
\label{zaehler}\\
&=&
9{n_g}^2-12 n_g-12
+  (12+18 n_g ) {n_h^t}
\sin{2\theta_{\tilde{t}}} \mu_t/\sin\beta 
\nonumber\\
&&+
(9 n_g^2 -12 n_g+20 -18 {n_h^t}^2 \mu_t^2/\sin^2\beta)
\cos{2\theta_{\tilde{t}}}
\label{zaehler2}
\eeqn
\beqn
{\cal F}^{D_1}&=&{I_{3L}^t}^2(n_g-2)^2+2 e_t I_{3L}^t(n_g-2)+2 e_t^2
+Y_t^2 {n_h^t}^2+\sqrt{2} \sin{2\theta_{\tilde{t}}} Y_tI_{3L}^t n_h^t (n_g-2)
\nonumber\\&&
+\cos{2\theta_{\tilde{t}}}[{I_{3L}^t}^2(n_g-2)^2+2 e_t I_{3L}^t (n_g-2)]
\label{nenner}\\
&=&9 n_g^2 -12 n_g+20
+18 {n_h^t}^2 \mu_t^2/\sin\beta^2+
18 n_h^t (n_g-2) \sin{2\theta_{\tilde{t}}}\mu_t/\sin\beta 
\nonumber\\&&
+(9 n_g^2-12 n_g-12)\cos{2\theta_{\tilde{t}}}
\label{nenner1a}\\
{\cal F}^{D_2}&=&\sqrt{2} Y_t I_{3L}^t n_h^t(n_g-2)
-\sin{2\theta_{\tilde{t}}}[2 e_t^2+2 e_t I_{3L}^t(n_g-2)-Y_t^2 {n_h^t}^2]
\nonumber\\&&
+\sqrt{2} \cos{2\theta_{\tilde{t}}} Y_t n_h^t[2 e_t+I_{3L}^t(n_g-2)]
\label{nenner2}\\
&=&18 n_h^t(n_g-2)\mu_t/\sin\beta
-(24 n_g-16 -18 {n_h^t}^2 \mu_t^2/\sin^2\beta)
\sin{2\theta_{\tilde{t}}}
\nonumber\\&&
+n_h^t(12+18 n_g)\cos{2\theta_{\tilde{t}}} \mu_t/\sin\beta 
\label{nenner2a}
\eeqn
The abbrevations $n_g$, $n_h^t$ for the gaugino and higgsino 
components are given by
\beqn
n_g&=& 1+ \cot\theta_W
\frac{N_{k2}}{N_{k1}}\ ,
\label{lepcoup}\\
n_h^t&=&\cot\theta_W\frac{N_{k4}}{N_{k1}} \ .
\label{yuk_up}
\eeqn
The expression for the $t$ polarisation in the $\tilde{t}_2$ decay 
can be derived from eq.~(\ref{appd_5a}) by  
changing the sign of all terms $\sim \sin2\theta_{\tilde{t}}$ 
and $\sim \cos 2 \theta_{\tilde{t}}$ in
eqs.~(\ref{zaehler})--(\ref{nenner2a}) .\\[.5em] 

\noindent
{\bf ii) {\boldmath $\tilde{b}\to t$} Transition}\\[1ex]
With the appropriate couplings inserted, the polarisation of the top quark 
in the decays $\tilde{b}_i\to t \tilde{\chi}^{\pm}_k$ 
can be written equivalently to eq.~(\ref{appd_5a}):
\beqn
P_{\tilde{b}_1\to t\tilde{\chi}^{\pm}_k}&=&
\frac{{\cal G}^{N} f_1}
{{\cal G}^{D_1}-{\cal G}^{D_2} f_2 } \ . 
\label{appd_neu}
\eeqn
The coefficients of the numerator ${\cal G}^N$ and denominator
${\cal G}^{D_1}$ and ${\cal G}^{D_2}$ are given by
\beqn
{\cal G}^N&=&-{c_h^+}^2\mu_t^2/\sin^2\beta+{c_h^-}^2 
\mu_b^2/\cos^2\beta +2
-( 2 \sqrt{2}  c_h^- \mu_b/\cos\beta )\sin2\theta_{\tilde{b}} \nonumber\\
&&-({c^+_h}^2 \mu_t^2/\sin^2\beta+{c^-_h}^2 \mu_b^2/\cos^2\beta-2)
\cos 2\theta_{\tilde{b}}\ , \label{char_num}\\
{\cal G}^{D_1}&=&{c_h^+}^2\mu_t^2/\sin^2\beta+{c_h^-}^2 
\mu_b^2/\cos^2\beta +2
-( 2 \sqrt{2}  c_h^- \mu_b/\cos\beta )\sin2\theta_{\tilde{b}} \nonumber\\
&&+({c^+_h}^2 \mu_t^2/\sin^2\beta-{c^-_h}^2 \mu_b^2/\cos^2\beta+2)
\cos 2\theta_{\tilde{b}} \ , \label{char_den1}\\
{\cal G}^{D_2}&=&-c_h^+(1+\cos2\theta_{\tilde{b}})
2 \sqrt{2}\mu_t/\sin\beta
+c_h^+ c_h^- 4 \sin 2 \theta_{\tilde{b}} \mu_t\mu_b/\sin 2\beta \ ,
\label{char_den2}
\eeqn
where  $c_h^+$ and $c_h^-$
are ratios of $U$ and $V$ mixing-matrix elements,
\beqn 
c_h^{+}&=&V_{k2}/U_{k1} \ , \label{char-c-1a}\\
c_h^{-}&=&U_{k2}/U_{k1} \ . \label{char-c-1b}
\eeqn
The components of the chargino mixing matrix in the high $\tan\beta$ 
approximation read:
\begin{eqnarray}
U_{12} & = & \phantom{+}U_{21}= 
\sqrt{W+M_2^2-\mu^2+2m_W^2}/\sqrt{2W} \ ,\\
U_{11} & = &-U_{22}= -{\rm sign}(\mu)
\sqrt{W-M_2^2-\mu^2+2m_W^2}/\sqrt{2W} \ ,\\
V_{12} & = & -V_{21}=-{\rm sign}(M_2)
\sqrt{W+M_2^2-\mu^2-2m_W^2}/\sqrt{2 W} \ ,\\
V_{11} & = & \phantom{+}V_{22}=
\sqrt{W-M_2^2-\mu^2-2m_W^2}/\sqrt{2 W}
\hspace*{.5cm} \label{1T}
\end{eqnarray}
with $W = \sqrt{(M_2^2+\mu^2+2m_W^2)^2-4 M_2^2\mu^2}$. 
The explicit expression for the $t$ polarisation in the decay
$\tilde{b}_2\to t \tilde{\chi}^{\pm}_k$ can be derived from
eq.~(\ref{appd_neu}) by changing in
eqs.~(\ref{char_num})--(\ref{char_den2}) the sign in the terms $\sim
\cos 2 \theta_{\tilde{b}}$ and $\sim \sin 2 \theta_{\tilde{b}}$.

\subsection{A Study of {\boldmath $t$} Polarisation}

Since the $t$ polarisation in 
the process $\tilde{t}_i\to t \tilde{\chi}^0_k$ depends on 
$1/\sin\beta$, eqs.~(\ref{zaehler})--(\ref{nenner2}), 
it is only weakly sensitive to large $\tan\beta$.
By contrast, the decay $\tilde{b}_1\to t \tilde{\chi}^{\pm}_1$ 
can be used indeed to measure $\tan\beta$.
A feasibility study of the reaction
\begin{eqnarray}
  e^+_Le^-_R & \to & \sb_1\bar \sb_1
  \ \to \ t\cx_1^\pm + \bar{t} \cx_k^\mp
  \quad  \ 
  \label{sb11}
\end{eqnarray}
has been performed at $\sqrt{s}=1.9~\TeV$ within the reference scenario $RP$. 
The cross section amounts to $\sigma_{\sb_1\sb_1} = 10~\fb$ assuming beam
polarisations of $P_{e^-} = +0.80$ and $P_{e^+} = -0.60$.

The top polarisation measurement requires the reconstruction of the
$t$ system and of the direction of the primary squark $\sb_1$.
If no other particles except the neutralinos escape detection and
all SUSY particle masses are known (as assumed in the present study),
it is possible to reconstruct the momenta of both $\nt_1$ kinematically.
The $\sb_1$ directions can then be determined up to a twofold ambiguity, 
where the correct solution gives the expected angular distribution
$\propto \sin\theta^2_{\tilde{b}_1}$ while the false solution contributes
uniformly in $\cos\theta_{\tilde{b}_1}$.
For a distribution measured with respect to the $\sb_1$ direction, 
like the strange quark in the top decay of eq.~(\ref{tdecay}), 
the ambiguity can be resolved on a statistical basis by subtracting the 
`wrong' solution ({\it e.g.} via a Monte Carlo simulation).
Therefore the following decay chains of reaction~(\ref{sb11})
have been considered:
\begin{eqnarray}
  \sb_1^{(1)} 
    & \to & \ \
     t \cx_1^\pm, \, \qquad t\to b\, W \to b\, c \bar s, \
                \quad \cx_1^\pm \to q \bar q'\, \nt_1 
                \qquad\qquad\quad\,  \cB^{(1)} = 0.076 \ ,
  \label{btc1}\\[1ex]
  \sb_1^{(2)} 
    & \to &
  \begin{cases}
     t \cx_1^\pm, \qquad t\to b\, W \to b\, q \bar q', 
                \quad \cx_1^\pm \to q \bar q'\, \nt_1 
                & \quad \cB^{(2)} = 0.152 \ , \\[1ex]
     t \cx_2^\pm, \qquad t\to b\, W \to b\, q \bar q', 
                \quad \cx_2^\pm \to q \bar q'\, (W / Z ) \,\nt_1 
                & \quad \cB^{(3)} = 0.180 \ ,
  \end{cases}    \label{btc12}
\end{eqnarray}
where the first sequence contains the decay of interest
and the $\cB^{(i)}$ denote the combined branching ratios.
The $\sb_1$ branching ratios to charginos are  
$\cB(\sb_1\to t \cx^\pm_1)=0.36$ and
$\cB(\sb_1\to t \cx^\pm_2)=0.30$.
The large number of combinatorics can be efficiently reduced by
requiring flavour identification 
-- two bottom jets from top decays and at least one charm jet from 
$W$ decays --
and applying additional kinematic constraints on the reconstruction
of $W$ masses, top masses and chargino $\cx_1^\pm$ or  $\cx_2^\pm$
masses.

The program {\sc CompHEP}~\cite{comphep} has been used to calculate
the exact decay distributions of the \mbox{5-particle} final state
$e^+e^-\to \sb_1 + t\,\cx_1^\pm \to \sb_1 + b c \bar{s}\,\cx_1^\pm$.
Flavour tagging efficiencies for bottom and charm jets of 
$\varepsilon_b=0.85$ and $\varepsilon_c=0.5$ 
with reasonable purities ($\sim 0.8$) have been assumed~\cite{TDR}.
With an integrated luminosity of $\cL=2,000~\fbi$ one expects
$N = 2\,\cB^{(1)}\,(\cB^{(2)}+\cB^{(3)})\,\sigma_{\sb_1\sb_1}
      \,\varepsilon\,\cL \simeq 330$ 
reconstructed $\sb_1\to t\cx_1^\pm$ decays.
The generated angular distribution $\cos\theta^*_s$, 
where $\theta^*_s$ is the angle between the 
$\bar{s}$ quark and the primary $\sb_1$ in the top rest frame,
is presented in Fig.~\ref{top-pol}.
A fit to the top polarisation, given by eq.~(\ref{tdecay}),
yields $P_t = -0.44 \pm 0.10$, which is consistent with the input
value of $P_t^{th} = -0.38$. 
From such a measurement one can derive $\tan\beta = 17.5 \pm 4.5$, 
as illustrated in Fig.~\ref{top-tb}.

\begin{figure}[htb] \centering
  \begin{picture}(11,8)
    \put(-1,0){\epsfig{file=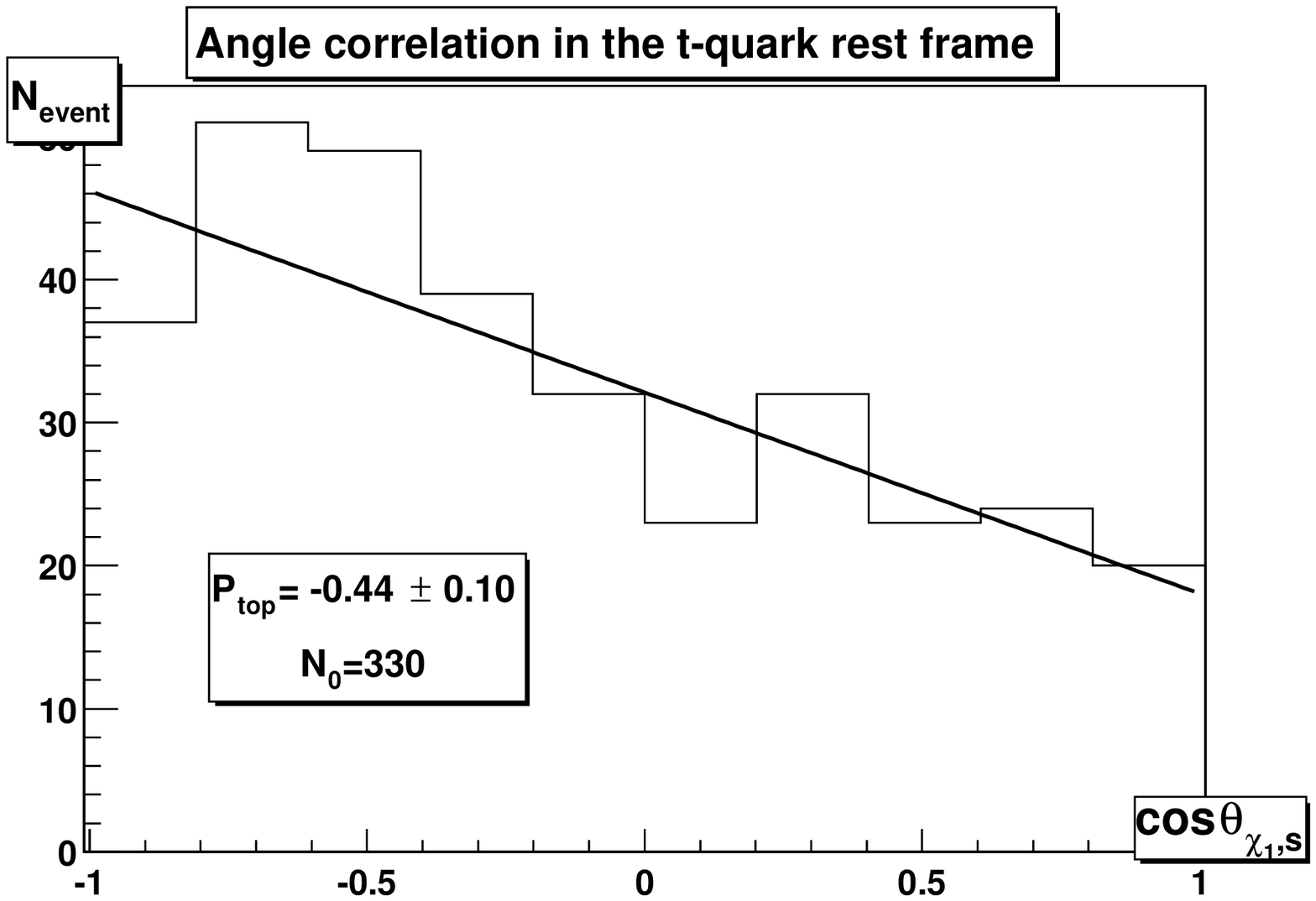,scale=.6}}
    \put(1,7.37){\epsfig{file=box.eps,width=150mm,height=10mm}}
    \put(1,1.3){\epsfig{file=box.eps,width=60mm,height=30mm}}
    \put(-.8,6.05){\epsfig{file=box.eps,width=10mm,height=10mm}}
  \end{picture}
  \caption{ \label{top-pol} \it 
    Angular distribution in $\cos\theta^*_s$, with $\theta^*_s$ 
    being the angle between the  $\sb_1$ 
    and $\bar s$ partons in the top rest frame of $t\to b\,c\bar s$
    decays from 
    $e^+_Le^-_R\to \sb_1\sb_1\to t\cx_1^\pm + \sb_1$ production
    at $\sqrt{s}=1.9~\TeV$. The histogram corresponds to
    $\cL=2,000~\fbi$, the line represents a fit to a top polarisation
    of $P_t = -0.44  \pm0.10$.}
\end{figure}

\begin{figure}[htb] \centering
  \begin{picture}(11,8)
    \put(-1,0.3){\epsfig{file=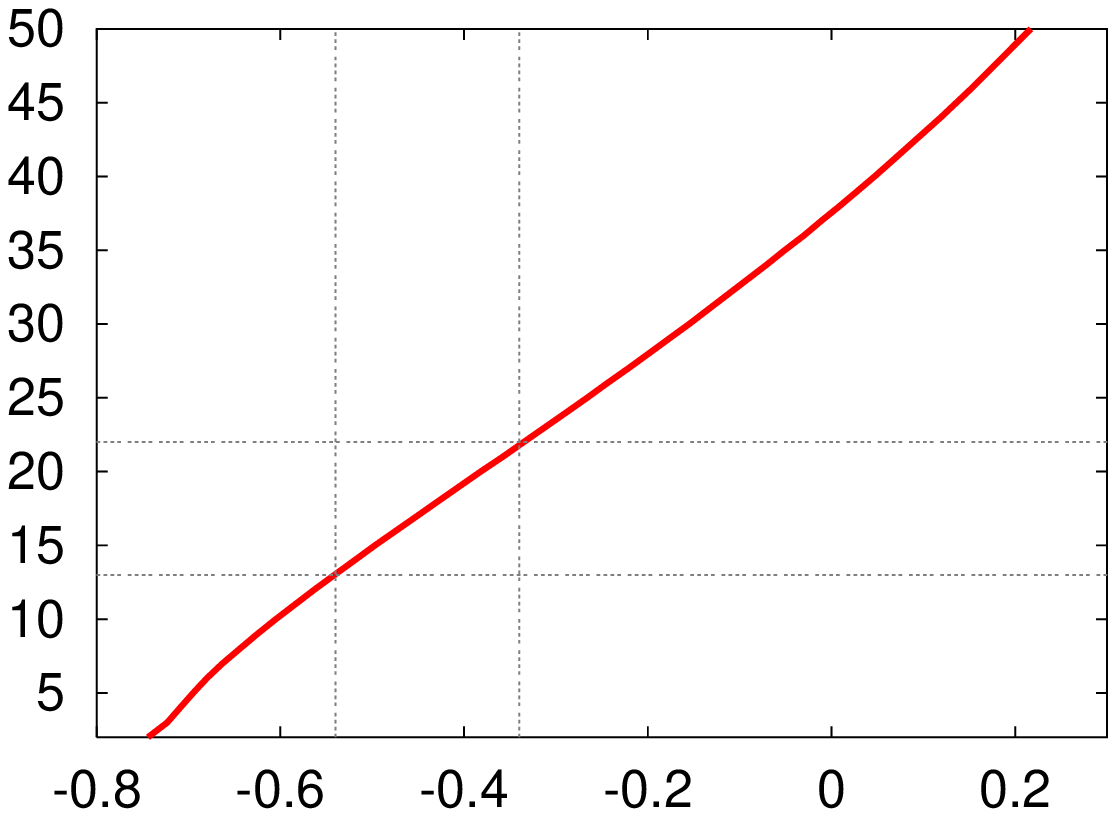,scale=.9}}
    \put(-2,6.8){\Large $\tan\beta$}
    \put(8.0,0){\Large $P_{\sb_1\to t\cx_1^\pm}$}
  \end{picture}
  \caption{ \it 
    $\tan\beta$ as a function of top polarisation 
    $P_{\sb_1\to t \cx_1^\pm}$ for reference scenario $RP$.
    The bands indicate the results of a simulation 
    with $P_t = -0.44\pm 0.10$ leading to $\tan\beta=17.5\pm4.5$. 
    \label{top-tb} }
\end{figure}

Using eq.~(\ref{offdiag_b}), the trilinear bottom coupling 
can be expressed as
\begin{eqnarray}
  A_b & = & 
   \frac{m_{\sb_1}+m_{\sb_2}}{2} \cdot
   \frac{m_{\sb_1}-m_{\sb_2}}{m_{b}} 
   \, \sin 2 \theta_{\sb} 
   + \mu \tan\beta \ ,
  \label{abottom} 
\end{eqnarray}
with a nominal value of $A_b = -778~\GeV$ in the scenario $RP$.
The uncertainty coming from the second term can be
considerably reduced when using the $\tan\beta$ measurement from the 
$\tau$ sector. 
The contribution to  $A_b$ amounts to 
$\delta A_b^{(\delta \tan\beta )} = 280~\GeV$.
The first term of eq.~(\ref{abottom}) requires a knowledge of this $\sb$ 
masses and mixing angle. 
From a cross section measurement of reaction (\ref{sb11}) 
with $\cL=2000~\fbi$ one expects 
for the mixing angle a statistical accuracy of 
$\sin 2 \theta_{\sb} = 0.82\pm0.04$, 
which corresponds to a contribution of
$\delta A_b^{(\delta \sin 2\theta_{\tilde{b}})} = 140~\GeV$.
Due to the small mass difference $m_{\sb_1}-m_{\sb_2} = 32~\GeV$
the precision on $A_b$ is limited by the errors on the masses.
Assuming $\delta m_{\sb}= 5~\GeV$ one gets
$\delta A_b^{(\delta m_{\sb})} = 770~\GeV$, which is of the same magnitude
as the trilinear coupling itself. 
If the mass determination can be improved 
to $\delta m_{\sb}= 2~\GeV$ or better, 
the trilinear coupling $A_b$ can be obtained with a 
statistical accuracy of $\delta(A_b)/A_b \sim 60\%$ or better.

The analysis can be carried out correspondingly 
for the trilinear top coupling:
\begin{eqnarray}
  A_t & = & 
   \frac{m_{\st_1}+m_{\st_2}}{2} \cdot
   \frac{m_{\st_1}-m_{\st_2}}{m_{t}} 
   \, \sin 2 \theta_{\st} 
   + \mu \cot\beta \ .
  \label{atop} 
\end{eqnarray}
In the reference scenario $RP$ the nominal value 
is given by $A_t=-510~\GeV$.
Due to the large value of $\tan\beta$ the second term in eq.~(\ref{atop})
is completely negligible.
Thus one relies solely on the mass and cross section measurements.
Assuming $\delta m_{\st}\simeq 10~\GeV$ and 
$\delta\sigma / \sigma = 0.05$, corresponding to
$\sin 2 \theta_{\st} = 0.92\pm0.06$, the top coupling can be
determined to an accuracy of $\delta(A_t)/A_t\lesssim 10\%$.

The above estimates of measurements of
the top polarisation from $\sb_1$ decays and the trilinear
bottom and top couplings can only serve as a rough guide. 
A more realistic statement must include background
from combinatorics and production of all other squarks. 
Such an analysis, however, can only be done with specific assumptions
on a detector performance and a jet reconstruction algorithm, which
goes beyond the scope of the present paper.

\section{Summary and Outlook}
%
\noindent
High-precision analyses of fundamental parameters 
will be crucial elements of high-energy physics in the future.
In supersymmetric theories they should allow us to bridge the gap 
from the electroweak scale to the grand unification / Planck 
scale in a stable way so as to set a link between particle physics 
and gravity.

The mixing angle $\tan{\beta}$ in the Higgs sector and the 
trilinear couplings $A_f$ in the soft SUSY breaking terms, correlated
with the interactions described by the superpotential, are parameters
in supersymmetric theories that are particularly difficult to 
determine. In this paper we have explored opportunities to measure
these parameters in pair production of stau, sbottom and stop
particles at prospective $e^+ e^-$ linear colliders. 

Analysing the $f$ polarisation in decays (generically) 
${\tilde f} \to f {\tilde{\chi}}$  
proves very promising for measurements of large values of $\tan{\beta}$
with an accuracy at the 10\% level. 

Measurements of the split sfermion
masses and the mixing angles can subsequently be exploited to
determine the trilinear couplings $A_f$. In some areas of the parameter
space in which the mass splitting is large and the mixing small,
it is very difficult to reach a precision beyond the order of 
magnitude, while in other areas the 10\% level can readily be 
achieved. 

A systematic screening of the parameter space will be
presented in a sequel to this report.

\section{Acknowledgements}
The authors thank G.~B\'elanger and S.~Kraml for useful discussions.
We are grateful to G.~A.~Blair and W.~Porod for the careful reading 
of the manuscript. E.B. and A.S. were partly supported by
the INTAS 00-0679 and CERN-INTAS 99-377 grants.
E.B. thanks the Humboldt Foundation for the Bessel Research Award
and DESY for the kind hospitality. G.M.-P. was partly supported by
the Graduate College `Zuk\"unftige Entwicklungen in der Teilchenphysik' 
at the University of Hamburg, Project No. GRK 602/1.
This work was also supported by the EU TMR Network Contract No.
HPRN-CT-2000-00149.

\begin{appendix}
 
\section{Analytical Expressions in the high {\boldmath $\tan\beta$}
Approximation \label{appA}}
In order to study the $\tan\beta$ dependence analytically,
we express the neutralino wave functions by
the MSSM parameters $M_{1,2}$ and $\mu$, which are 
given in compact form in Ref.~\ci{CKMZ}.
In the high $\tan\beta$ sector we can use the approximations
\begin{eqnarray}
\sin2\beta&\approx& 2 \tan^{-1}\beta \,\, (1-\tan^{-2}\beta)\approx
2 \tan^{-1}\beta,\label{sin-app}\\
\cos2\beta&\approx& 2\tan^{-2} \beta -1\approx -1 \, , \label{cos-app}
\end{eqnarray}
which lead for the gaugino and higgsino coefficients $n_g$ and $n_h$ 
in the ${\tilde{\tau}}_i \to \tau {\tilde\chi}_k^0$ decay 
to the expressions:
\begin{eqnarray}
n_g&=&2+\frac{1}{\sin\theta_W \cos\theta_W B_k/A_k-\sin^2\theta_W}
\label{eq_xw} \, \\
n_h&=&\frac{\tilde{C}_k/A_k+\tilde{D}_k}{\sin\theta_W
(B_k/A_k-\tan\theta_W)} \,
\label{eq_wh}
\end{eqnarray}
where 
\begin{eqnarray}
A_k&=&m_Z^2(M_2^2\sin^4\theta_W+M_1^2 \cos^4\theta_W+2 M_2 M_1
\sin^2\theta_W \cos^2\theta_W -m_k^2) \nonumber\\
&&+\, (M^2_2 \sin^2\theta_W+M_1^2 \cos^2\theta_W-m_k^2)(\mu^2-m_k^2)
\label{appb_eq1a}\\
B_k&=&\sin\theta_W \cos\theta_W [m_Z^2(M_1 \cos^2\theta_W+M_2
\sin^2\theta_W)(M_1-M_2)+2m_Z^2\mu(M_2-M_1)/\tan\beta\nonumber\\
&&-\,(\mu^2-m_k^2)(M_2^2-M_1^2)]\label{appb_eq1b}\\
\tilde{C}_k&=&m_Z[M_1 \sin^2\theta_W(m_k^2-M_2^2)+M_2 
\cos^2\theta_W(m_k^2-M_1^2)]/\tan\beta \label{appb_eq1c}\\
\tilde{D}_k&=&-\frac{m_Z \mu }{\mu^2-m_k^2} \ . \label{appb_eq1d} 
\end{eqnarray}
For completeness we note that the 
mass eigenvalues behave as $m_k^2=1+{\rm const}/{\tan\beta}$.
(The expressions $\tilde{C}_k$ and $\tilde{D}_k$ of
(\ref{appb_eq1c}) and (\ref{appb_eq1d}) 
correspond to $\cos\beta \, C_k$ and $\sin\beta \, D_k$ in the 
notation of Ref.~\cite{CKMZ}).

\section{Monte Carlo Study of {\boldmath $e^+e^-\to \stau_1\stau_1$} }
\label{mcstudy}

In order to get a more realistic estimate of the achievable precision
of the $\stau$ parameters a detailed simulation of the process
\begin{eqnarray} 
  e^+_L e^-_R & \to & \stau^+_1\stau^-_1 
              \ \to \ \tau^+ \nt_1 + \tau^- \nt_1 
  \label{st1prod}
\end{eqnarray} 
has been performed, assuming reference scenario $RP$ with
beam polarisations of 
$P_{e^-} = +0.80$ and $P_{e^+} = -0.60$,
a cm energy of $\sqrt{s}=500~\GeV$ and an integrated luminosity
of $\cL=250~\fbi$.
The SUSY particles masses are
$m_{\stau_1} = 154.6~\GeV$ and
$m_{\nt_1} = 78.0~\GeV$,
the $\stau_1$ decay modes and branching ratios are
$\cB(\stau_1\to\tau^-\nt_1) = 0.779$,
$\cB(\stau_1\to\tau^-\nt_2) = 0.124$ and
$\cB(\stau_1\to\nu_\tau\cx_1^-) = 0.097$.

Events are generated using the Monte Carlo program 
{\sc Pythia~6.2}~\cite{pythia}, which includes beam polarisation,
QED radiation and beamstrahlung effects~\cite{circe}.
Polarised $\tau$ decays are treated by an interface to
{\sc Tauola}~\cite{TAUOLA}.
The detector properties, acceptances and resolutions, follow
the concept described in 
\cite{TDR}
and realised in the parametric simulation program 
{\sc Simdet}~\cite{simdet}.

The $\tau$ identification proceeds via the leptonic decays
$\tau \to \ell \bar\nu_\ell \nu_\tau$ with $\ell = e$ or $\mu$,
and the hadronic decays $\tau \to h \,\nu_\tau$
with $h = \pi,\ \rho\,(\pi\pi^0)$ or generic
$3\,\pi\,(\pi\pi^+\pi^- + \pi\pi^0\pi^0)$ final states.
The experimental signature for reaction (\ref{st1prod})
are two acoplanar $\tau$ candidates, excluding di-lepton final states,
and large missing energy.
Background from Standard Model processes is suppressed by demanding
the $\tau$'s to carry less than the beam energy 
($E_{\ell,\,h} < 0.8\,E_{\rm beam}$),
to be produced in the central region 
(polar angle $|\cos \theta_\tau| < 0.75$)
and to be acoplanar (azimuthal angles $\Delta\phi < 160^\circ$).
Two-photon contributions $\ee\to\ee \tau^+ \tau^-$ are completely
rejected by vetoing scattered electrons and radiative photons 
down to polar angles $\theta > 4.6$~mrad.
The remaining background from $WW$ production is $\sim 6\%$.
Other contributions come from SUSY processes, in particular from chargino 
and neutralino production. 
The reaction 
$\ee\to\cx_1^+\cx_1^-\to \tau^+\nu\nt_1\,\tau^-\nu\nt_1$
has a similar topology, but a softer $\tau$ energy distribution,
and it amounts to $\sim 3\%$.
Neutralino production
$\ee\to\nt_2\nt_1\to\tau^+\tau^-\nt_1\nt_1$ is large.
The $\tau\tau$ pair tends to be more collinear, and
demanding an acollinearity angle $\xi>90^\circ$ suppresses
this contribution to a level of $\sim 7\%$.

Applying these event selection criteria to a complete simulation of
signal and background processes,
the experimental cross section, including QED radiation and beamstrahlung 
effects, can be determined as 
\begin{eqnarray} 
  \sigma (e^+_Le^-_R\to\stau_1\stau_1) & = &
  \frac{N_{\tau\tau} - N_{bkg}}{\varepsilon\cdot\cL} 
    \ = \ 113.5 \pm 1.4\,({\rm stat}) \pm 3.3\,({\rm sys})~\fb \ ,
  \label{xsection}
\end{eqnarray} 
where the first error represents the statistical and the second error
the systematic uncertainties. 
The overall efficiency is $\varepsilon \simeq 0.20\pm0.006$ and
includes the branching ratios $\cB_{\stau_1}$ and $\cB_\tau$.
Obviously, the expected precision is limited by systematics. The
dominant error comes from the $\stau_1$ branching ratio, which will be
difficult to measure with high accuracy. 
It is assumed that a value of $\cB(\stau_1\to\tau\nt_1)=0.78 \pm 0.01$
may be achieved finally, although higher rates are needed
than used in the present study.
Further systematics to be considered are the precise knowledge of
background, acceptance corrections, the degree of beam polarisations
and the $\tau$ decay rates. 
The sum of all sources gives an estimate of 
$\delta\sigma/\sigma \simeq 3\,\%$. 
Note that the result for the cross section 
of eq.~(\ref{xsection}) needs to be
corrected for radiative effects before comparing to the 
Born calculation given in Table~\ref{tab_rates}.

The $\stau$ mass can be determined from the shape of the hadronic
energy spectra of $\tau$ decays.
The isotropic two-body decay $\stau\to\tau\nt_1$ leads to a uniform
$\tau$ energy distribution in the laboratory system. 
The `endpoints' of the energy spectrum, in the usual notation
\begin{eqnarray} 
  E_{+/-} & = &
        \frac{m_{\stau}}{2} 
        \left ( 1 - \frac{m_{\cx}^2}{m_{\stau}^2} \right ) 
        \frac{1 \pm \beta}{\sqrt{1 - \beta^2}} \ ,
\end{eqnarray}
can be used to derive the masses of the primary $\stau$
\begin{eqnarray}
        m_{\stau} & = &
        \frac{\sqrt{E_{-} \cdot E_{+}}}
                       {E_{-}+E_{+}} \,\sqrt{s} \ ,
\end{eqnarray}
and the neutralino $\nt_1$ (assumed to be known in the present
analysis).
The resulting hadron spectra of $\tau$ decays are of triangular
shape -- modified by mass effects and detector resolution --
and they are still sensitive to $m_{\stau_1}$.
The energy distributions peak (turn over)
at the lower endpoint $E_-=20.0~\GeV$ and extend 
up to the upper endpoint $E_+=166.4~\GeV$,
see Figs.~\ref{st1mass} and \ref{st1pol}. 
On the other hand the shape of the energy spectrum 
also depends on the
$\tau$ polarisation, as discussed above for the $\pi$ spectrum.
Fortunately the $P_\tau$ dependence is very weak for the $\rho$
spectrum  and essentially absent for the $3\pi$ final states 
and there is no need for a two-parameter analysis in these channels.
The simulated energy spectra $E_\rho$ and $E_{3\pi}$ are shown
in Fig.~\ref{st1mass}.
A fit to the $\rho$ spectrum yields a $\stau_1$ mass of
$m_{\stau_1}=155.2 \pm 0.8~\GeV$.
The analysis of the $3\pi$ spectrum gives a slightly better 
resolution with $m_{\stau_1}=154.8 \pm 0.5~\GeV$.
Both results are consistent with the nominal value of $154.6~\GeV$.

\begin{figure}
  \hspace{-3mm}
  \epsfig{file=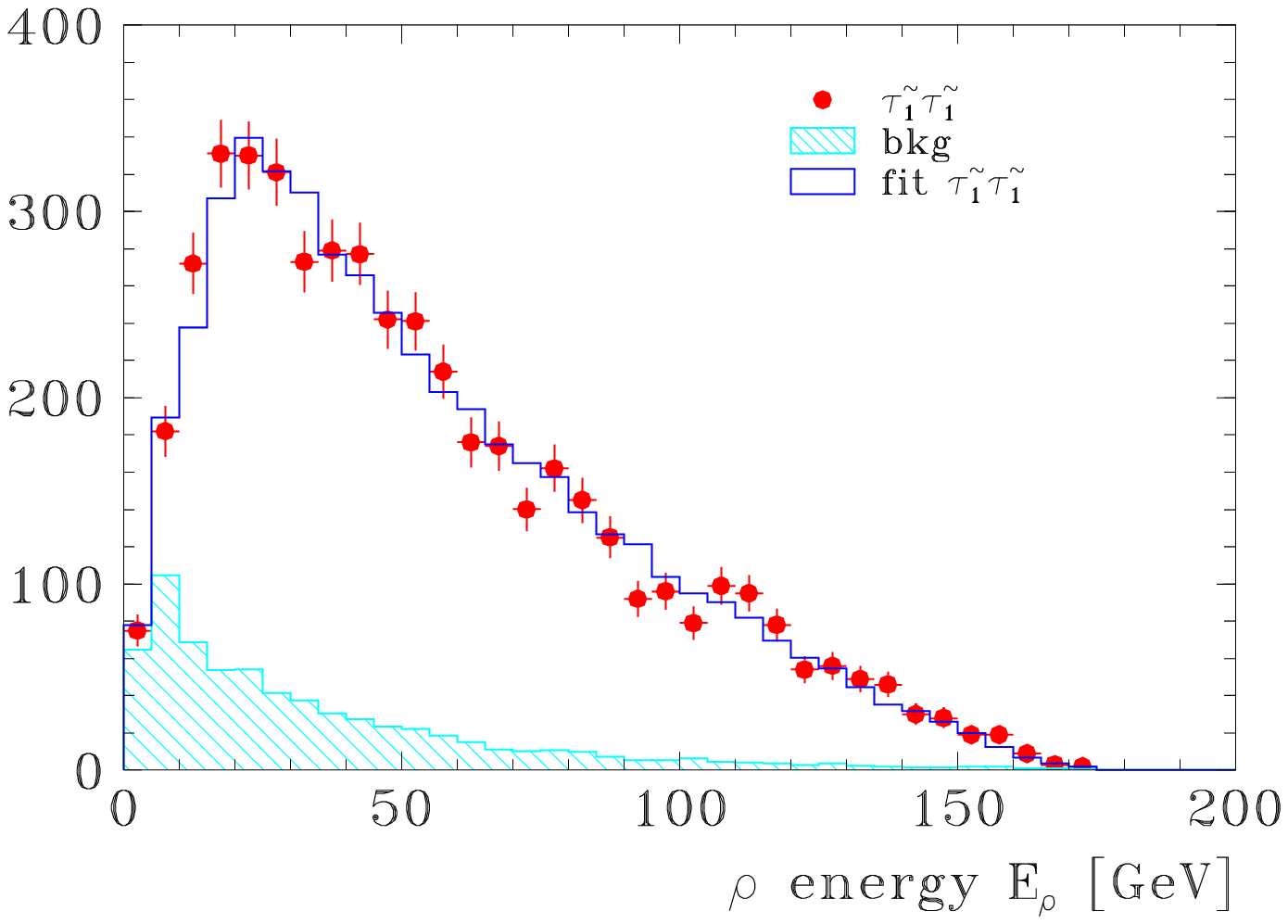,width=.5\textwidth}
  \epsfig{file=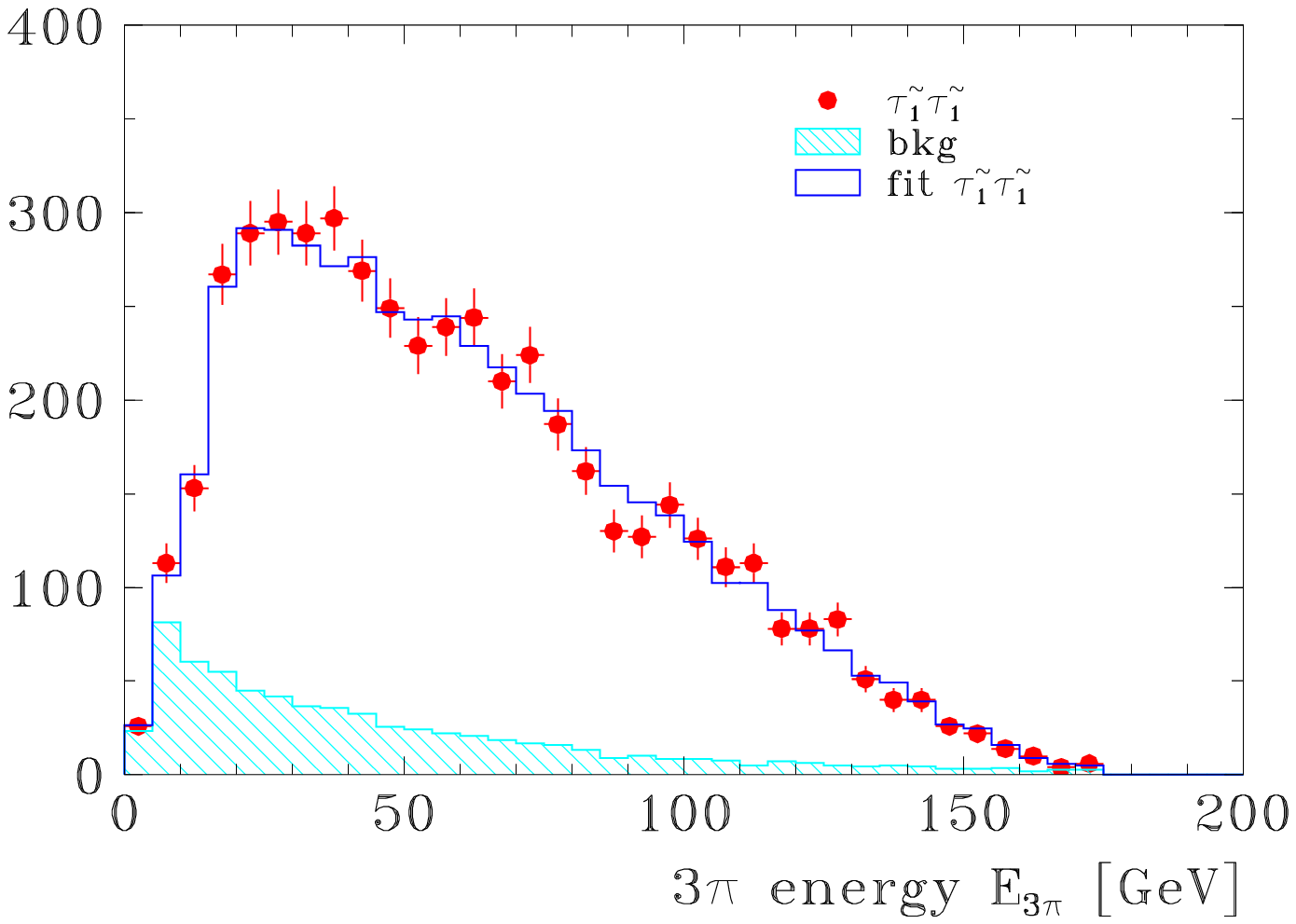,width=.5\textwidth}
  \caption{{\it 
      Hadron energy spectra of $\tau \to \rho \,\nu_\tau$
      and $\tau \to 3\pi \,\nu_\tau$ decays from 
      $e^+_L e^-_R \to \stau^+_1\stau^-_1$ production 
      at $\sqrt{s} = 500 ~\GeV$ with $P_{e^-}=+80\%$, $P_{e^+}=-60\%$
      assuming $\cL=250~\fbi$.
      The simulated data (dots) including SM and SUSY
      background (shaded histogram) are shown together with
      a fit to the $\stau_1$ mass.}}
  \label{st1mass}
\end{figure}
\begin{figure}[h]
  \hspace{-3mm}
  \epsfig{file=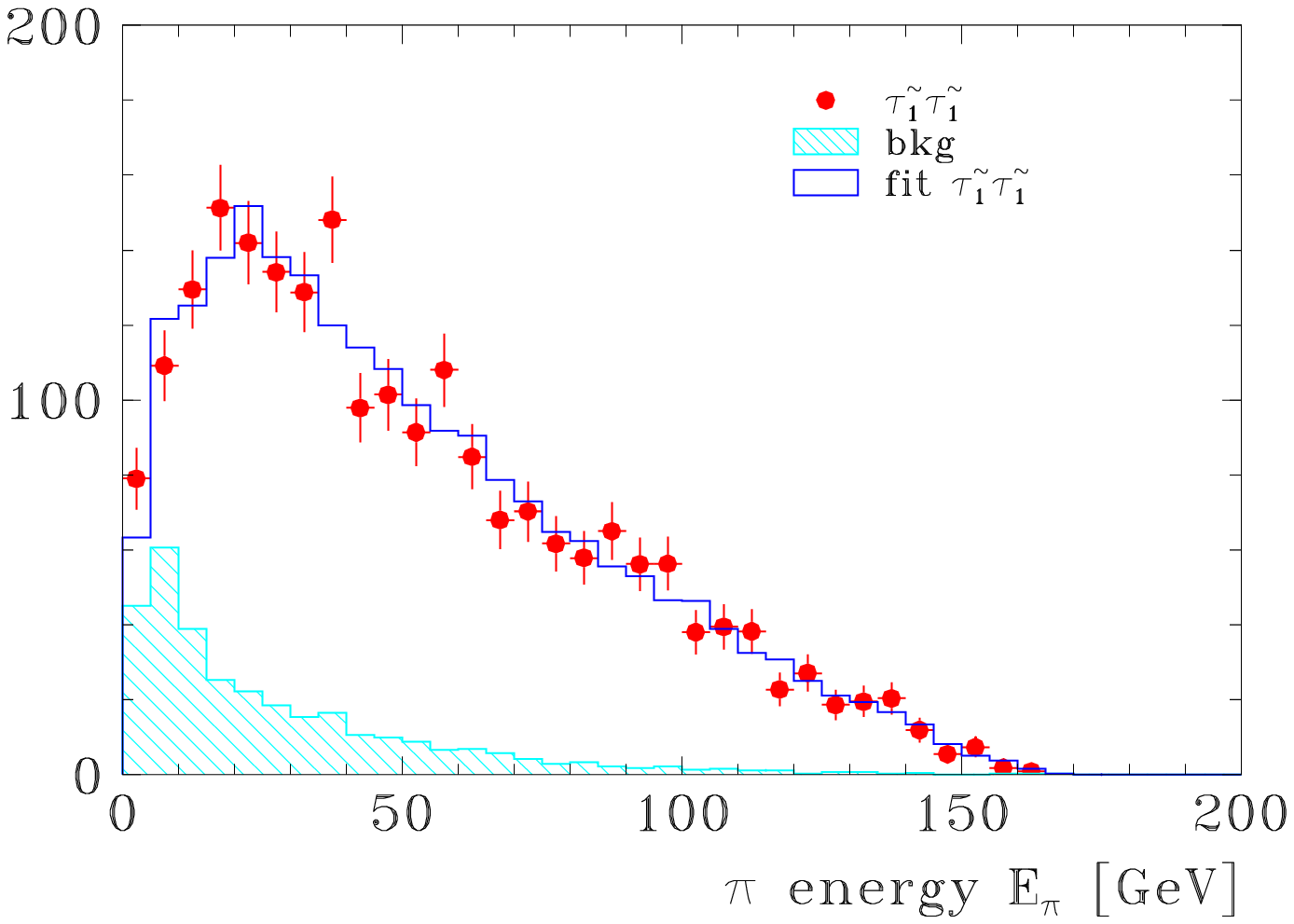,width=.5\textwidth}
  \epsfig{file=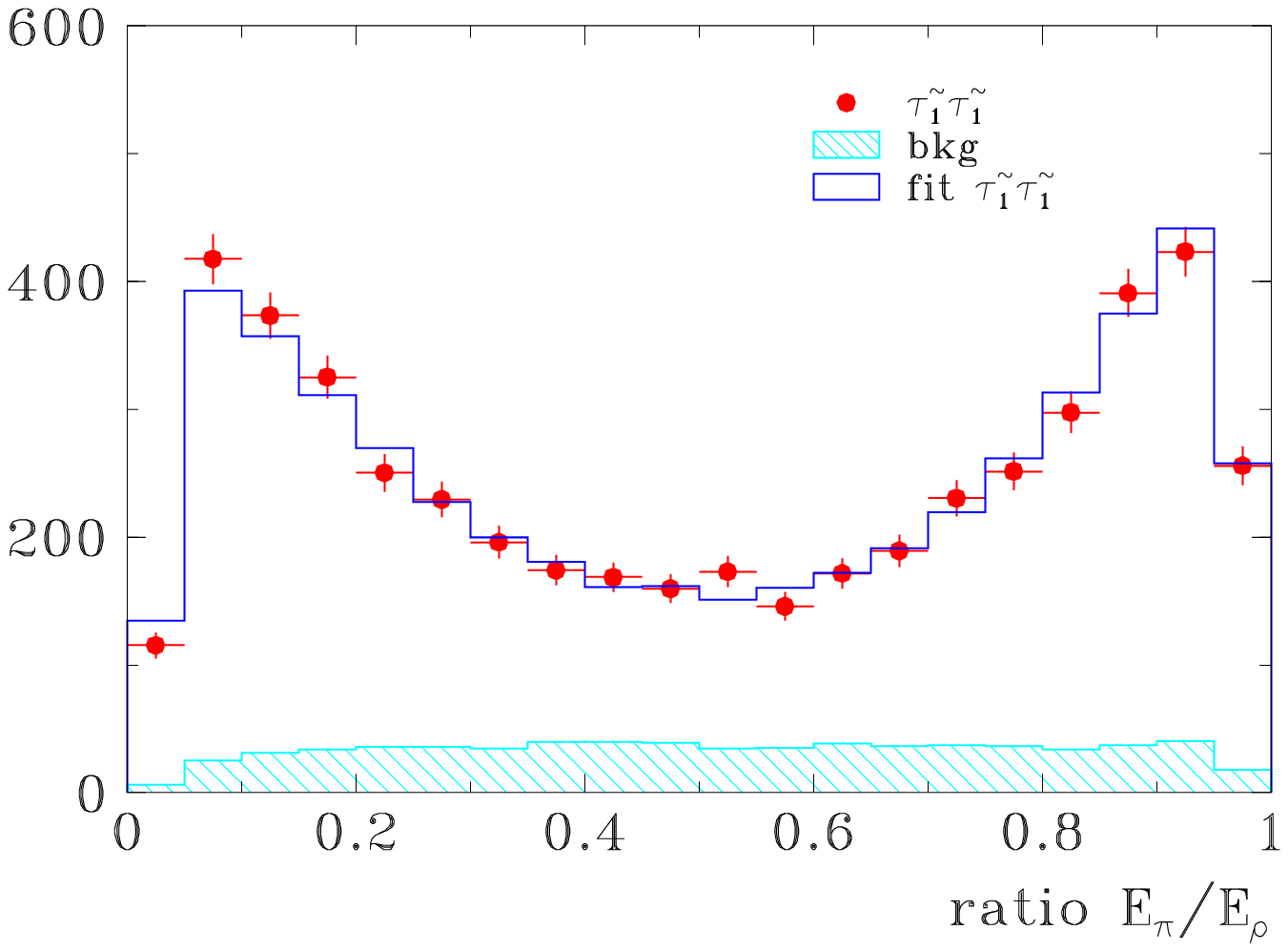,width=.5\textwidth}
  \caption{{\it 
      Pion energy spectrum of $\tau \to \pi \,\nu_\tau$
      and ratio $E_\pi/E_\rho$ of
      $\tau \to \rho \,\nu_\tau \to \pi\pi^0\,\nu_\tau $
      decays from 
      $e^+_L e^-_R \to \stau^+_1\stau^-_1$ production 
      at $\sqrt{s} = 500 ~\GeV$ assuming $\cL=250~\fbi$.
      The simulated data (dots) including SM and SUSY
      background (shaded histogram) are shown together with
      a fit to the $\tau$ polarisation $\cP_\tau$ from $\stau_1$ decays.}}
  \label{st1pol}
\end{figure}

Knowing the $\stau_1$ and $\nt_1$ masses, the $\pi$ energy spectrum and 
the decay characteristics of $\rho\to\pi\pi^0$ can be used to
determine the $\tau$ polarisation.
The energy distribution is harder for a $\pi$ emitted from a 
right-handed $\tau_R$ than from a left-handed $\tau_L$, as 
discussed above.
The $E_\pi$ spectrum is shown in Fig.~\ref{st1pol}.
A fit yields a $\tau$ polarisation of $P_\tau = 0.860 \pm 0.050$,
consistent with the input value of $P_\tau^{th} =0.85$.
Note that the residual background at low energies slightly reduces
the sensitivity.
In contrast to the $E_\rho$ distribution, there is a strong
$P_\tau$ dependence on the $\rho$ polarisation, which can be 
measured through the decay $\rho\to \pi\pi^0$.
Define the fraction of the energy carried by the charged $\pi$
as $z_\pi= E_\pi/E_\rho$.
A right-handed $\tau_R$ prefers to decay into a longitudinally 
polarised $\rho_L$, and the $z_\pi$ distribution 
$\sim (2 z_\pi-1)^2$ is peaked at $z_\pi\to 0$ and $1$,
{\it i.e.} most of the energy is carried by one of the pions.
A left-handed $\tau_L$ decays dominantly into a transversely polarised
$\rho_T$, resulting in a $z_\pi$ distribution 
$\sim 2 z_\pi\,(1 - z_\pi)$, {\it i.e.} a rather equal sharing of the 
energy between the two pions.
The distribution of the ratio $E_\pi/E_\rho$ is shown in
Fig.~\ref{st1pol}, 
with a flat contribution from  the unpolarised background.
From a fit to the $z_\pi$ distribution a value of
$P_\tau = 0.859 \pm 0.045$ is obtained.

{\it In summary.} The simulation of the reaction
$e^+_L e^-_R \to \stau^+_1\stau^-_1$ 
with $\cL=250~\fbi$ at $\sqrt{s} = 500 ~\GeV$
and beam polarisations of $P_{e^-}=+0.8$ and $P_{e^+}=-0.6$
demonstrates that the $\stau$ parameters can be determined with high 
precision.
The cross section measurement appears feasible with an accuracy
of $\delta\sigma/\sigma \simeq 3\,\%$, where the error is entirely due
to systematics and is dominated by the uncertainty of the branching ratio.
Using the information from all decay channels,
the $\stau_1$ mass can be determined with
an error of $\delta m_{\stau_1}=0.5~\GeV$ and 
the $\tau$ polarisation from $\stau_1$ decays can be measured
with an accuracy of $\delta P_\tau = 0.035$.
As shown in Fig.~\ref{sigpol}, the cross section depends sensitively
on the $\stau$ mixing angle.
Applying eq.~(\ref{App-59}) and including the experimental resolutions,
a value of 
$\cos 2\theta_\stau=-0.987\pm 0.02\,({\rm stat}) \pm 0.06\,({\rm sys})$ 
can be derived for the $\stau$ mixing angle.

\end{appendix}


\begin{thebibliography}{99}
\bibitem{All1} 
  J.~Wess and B.~Zumino,
  Nucl.\ Phys.\ B {\bf 70} (1974) 39.
\bibitem{All2}
  H.~P.~Nilles,
  Phys.\ Rept.\  {\bf 110} (1984) 1;
  H.~E.~Haber and G.~L.~Kane, Phys. Rep. {\bf 117} (1985) 75. 
 \bibitem{Extrapolations}
   G.~A.~Blair, W.~Porod and P.~M.~Zerwas,
   Phys. Rev. D {\bf 63} (2001) 017703 [hep-ph/0007107];
   G.~A.~Blair, W.~Porod and P.~M.~Zerwas, Eur. Phys. J. C {\it in press}
   [hep-ph/0210058].
 \bibitem{LC} E.~Accomando {\it et al.}  
   [ECFA/DESY LC Physics Working Group Collaboration],
   Phys.\ Rept.\  {\bf 299} (1998) 1 [hep-ph/9705442].
\bibitem{TDR}
J.~A.~Aguilar-Saavedra {\it et al.}, 
  ``TESLA Technical Design Report, 
  Part III: Physics at an e+e- Linear Collider,
  Part IV: A Detector for TESLA'',
  DESY 2001-011 [hep-ph/0106315].
 \bibitem{All-Parameters} G.~Moortgat-Pick {\it et al.},
hep-ph/0210212;
   A.~Freitas {\it et al.},
hep-ph/0211108;
   P.~M.~Zerwas {\it et al.},
hep-ph/0211076;
to appear in the Proceedings of
the 31st International Conference on High Energy
Physics (ICHEP 2002), Amsterdam, 2002.
 \bibitem{Choi}
   C.~Bl\"ochinger, H.~Fraas, G.~Moortgat-Pick and W.~Porod,
   Eur.\ Phys.\ J.\ C {\bf 24} (2002) 297 [hep-ph/0201282];
   S.~Y.~Choi, A.~Djouadi, M.~Guchait, J.~Kalinowski, H.~S.~Song and
   P.~M.~Zerwas,
   Eur.\ Phys.\ J.\ C {\bf 14} (2000) 535 [hep-ph/0002033].
   G.~Moortgat-Pick, A.~Bartl, H.~Fraas and W.~Majerotto,
   Eur.\ Phys.\ J.\ C {\bf 18} (2000) 379 [hep-ph/0007222];
   J.~L.~Kneur and G.~Moultaka,
   Phys.\ Rev.\ D {\bf 59} (1999) 015005 [hep-ph/9807336];
   G.~Moortgat-Pick and H.~Fraas,
   Acta Phys.\ Polon.\ B {\bf 30} (1999) 1999 [hep-ph/9904209].
\bibitem{CKMZ} S.~Y.~Choi, J.~Kalinowski, G.~Moortgat-Pick and P.~M.~Zerwas,
  Eur.\ Phys.\ J.\ C {\bf 22} (2001) 563 [hep-ph/0108117];
  S.~Y.~Choi, J.~Kalinowski, G.~Moortgat-Pick and P.~M.~Zerwas,
  Eur.\ Phys.\ J.\ C {\bf 22} (2001) 769 [hep-ph/0202039].
\bibitem{nachtrag} 
  H.~Baer, C.~H.~Chen, M.~Drees, F.~Paige and X.~Tata,
  Phys.\ Rev.\ D {\bf 59} (1999) 055014 [hep-ph/9809223]; 
  J.~L.~Feng and T.~Moroi,
  Nucl.\ Phys.\ Proc.\ Suppl.\  {\bf 62} (1998) 108 [hep-ph/9707494];
  V.~D.~Barger, T.~Han and J.~Jiang,
  Phys.\ Rev.\ D {\bf 63} (2001) 075002 [hep-ph/0006223]; 
  J.~F.~Gunion, T.~Han, J.~Jiang and A.~Sopczak, hep-ph/0212151.
\bibitem{susy02}
E.~Boos, G.~Moortgat-Pick, H.~U.~Martyn, M.~Sachwitz and A.~Vologdin,
hep-ph/0211040 and to appear in the 
Proceedings of the 10th International Conference on Supersymmetry 
and Unification of Fundamental Interactions, SUSY02, DESY, Hamburg 2002.
\bibitem{Noji} M.~M.~Nojiri,
  Phys.\ Rev.\ D {\bf 51} (1995) 6281 [hep-ph/9412374]; 
  M.~M.~Nojiri, K.~Fujii and T.~Tsukamoto,
  Phys.\ Rev.\ D {\bf 54} (1996) 6756 [hep-ph/9606370].
\bibitem{SPS} B.~C.~Allanach {\it et al.},
  Eur. Phys. J.C {\bf 25} (2002) 113 [hep-ph/0202233];
  N.~Ghodbane and \mbox{H.-U.~Martyn},
  hep-ph/0201233.
\bibitem{stau} 
S.~Kraml, PhD Thesis, HEPHY Vienna,
[hep-ph/9903257];
  A.~Bartl, H.~Eberl, S.~Kraml, W.~Majerotto and W.~Porod,
  Eur.\ Phys.\ J.\ directC {\bf 2} (2000);
  A.~Bartl, K.~Hidaka, T.~Kernreiter and W.~Porod, hep-ph/0207186;
  A.~Bartl, K.~Hidaka, T.~Kernreiter and W.~Porod,
  Phys.\ Lett.\ B {\bf 538} (2002) 137 [hep-ph/0204071];
  M.~Guchait and D.~P.~Roy,
  Phys.\ Lett.\ B {\bf 535} (2002) 243 [hep-ph/0205015].
\bibitem{Bartl:1989ms}
  A.~Bartl, H.~Fraas, W.~Majerotto and N.~Oshimo,
  Phys.\ Rev.\ D {\bf 40} (1989) 1594; 
  G.~Moortgat-Pick, A.~Bartl, H.~Fraas and W.~Majerotto,
  LC-TH-2000-032, hep-ph/0002253.
\bibitem{Microomegas} G. B\'elanger, F. Boudjema, A. Pukhov and A. Semenov,
  Comp. Phys. Comm. {\bf 149} (2002) 103.
\bibitem{comphep} A.~Pukhov, E.~Boos, M.~Dubinin, V.~Ilyin, D.~Kovalenko,
  A.~Kryukov, V.~Savrin, S.~Shichanin and A.~Semenov, Report INP-MSU 98-41/542,
  hep-ph/9908288; 
  A.~Semenov, Comp. Phys. Comm. {\bf 115} (1998) 124 and hep-ph/0205020.
\bibitem{TAUOLA}  S.~Jadach, Z.~Was, R.~Decker and J.~H.~K\"uhn,
  Comp. Phys. Comm. {\bf 76} (1993) 361.
\bibitem{boos2}
M.~Jezabek and J.~H.~K\"uhn, Nucl.\ Phys.\ B {\bf 320}
              (1989) 20;
 E.~E.~Boos and A.~V.~Sherstnev,
  Phys.\ Lett.\ B {\bf 534} (2002) 97
  [hep-ph/0201271].
\bibitem{pythia} T.~Sj\"ostrand {\it et al.}, 
  Comput. Phys. Commun. {\bf 135} (2001) 238 [hep-ph/0108264].
\bibitem{circe} T.~Ohl, 
  Comput. Phys. Commun. {\bf 101} (1997) 269 [hep-ph/9607454].
\bibitem{simdet} M.~Pohl and H.~J.~Schreiber, 
  DESY-02-061 and hep-ex/0206009.

\end{thebibliography}
\end{document}